\documentclass[11pt]{article}
\baselineskip
\usepackage{amssymb}
\usepackage{graphicx}
\usepackage{latexsym,cite}
\newcommand{\eq}{\begin{equation}}
\newcommand{\en}{\end{equation}}
\newcommand{\qe}{\end{equation}}
\newcommand{\ear}{\begin{eqnarray}}
\newcommand{\eqa}{\begin{eqnarray}}
\newcommand{\rae}{\end{eqnarray}}
\newcommand{\ena}{\end{eqnarray}}
\newcommand{\beq}{\begin{equation}}
\newcommand{\eeq}{\end{equation}}
\newcommand{\bea}{\begin{eqnarray}}
\newcommand{\eea}{\end{eqnarray}}
\newcommand{\Z}{\mathbb{Z}}
\newcommand{\N}{\mathbb{N}}
\newcommand{\bra}{\langle}
\newcommand{\ket}{\rangle}
\newcommand{\tr}{\mbox{tr}} 

\begin{document}
\begin{titlepage}
\vskip0.5cm
\begin{flushright}
DIAS-STP-05-04\\
\end{flushright}
\vskip0.5cm
\begin{center}
{\Large\bf  A numerical study of confinement in compact QED}
\end{center}
\vskip1.3cm
\centerline{Marco Panero}
 \vskip1.0cm
 \centerline{\sl  School of Theoretical Physics,
Dublin Institute for Advanced Studies,}
 \centerline{\sl  10 Burlington Road, Dublin 4, Ireland}
 \centerline{\sl e-mail: \hskip 1cm  panero@stp.dias.ie}
 \vskip1.0cm

\begin{abstract}
Compact U(1) lattice gauge theory in four dimensions is studied by means of an efficient algorithm which exploits the duality transformation properties of the model. We focus our attention onto the confining regime, considering the interquark potential and force, and the electric field induced by two infinitely heavy sources. We consider both the zero and finite temperature setting, and compare the theoretical predictions derived from the effective string model and the dual superconductor scenario to the numerical results.
\end{abstract}
\end{titlepage}

\section{Introduction}\label{introsect}

Confinement of the color degrees of freedom in chromo-singlet hadronic states is one of the most intriguing phenomena taking place in the low energy regime of quantum gauge theories. The problem cannot be studied perturbatively, but a number of different tools have been used, in order to find a theoretical description of confinement; in the years, several different objects have been proposed as responsible for the mechanism driving this phenomenon --- see \cite{Greensite:2003bk,Engelhardt:2004pf} for a review --- and, although a complete analytical solution is still missing, considerable progress has been achieved towards the theoretical interpretation of this problem \cite{Confinement2003}. 

Regardless of the non-universal details in their microscopic dynamics, different gauge theories share common properties in the infrared regime of the confined phase, where some effective degrees of freedom may account for the observed phenomena. 

In particular, according to the effective string scenario \cite{Luscher:1980fr,Luscher:1980iy,Luscher:1980ac}, the long distance properties of the potential of a confined charge-anticharge pair ($Q\bar{Q}$) in the heavy quark limit are expected to be described in terms of a fluctuating string; as it will be discussed in the following, some predictions of this picture have been successfully confirmed by numerical tests, but other aspects are still to be understood. Also, lattice results seem to indicate that the fine details could be different for different confining gauge theories.

At present, the details about a possible effective theory to provide a description of gauge models in the IR regime are not completely known yet, and a closer look at the behaviour of different gauge models may be useful to get some insight into the confinement dynamics.

In this paper, we focus our attention onto compact U(1) lattice gauge model in $D=4$: an abelian theory which is known to possess a confined phase, analogous to non-abelian models. We restrict our attention to the infinite quark mass limit, considering a system with a static $Q\bar{Q}$ pair of external charges, and work out numerical results to be compared with the theoretical expectations for the interquark potential $V(r)$ and for the force $F(r)$. We study both the zero and finite temperature cases. For this model, the values of the transition critical coupling at different finite temperatures have been published in a recent study \cite{Vettorazzo:2004cr}, and for these values we focus our attention onto different ratios $T/T_c$, in order to point out the quantitative effects that show up at finite temperature.

Then, we study the electric field induced by the $Q\bar{Q}$ pair, measuring the flux-tube profile in the middle plane between the two charges: according to the ``dual superconductor'' hypothesis, in the confined phase the flux lines joining the sources are expected to get ``squeezed'' in a tube. Similar studies about this model can be found in the literature --- see, for instance, \cite{Zach:1995ni,Zach:1997yz,Koma:2003gi}.

The numerical results are obtained by means of a Monte Carlo algorithm which exploits an exact, analytical mapping of U(1) gauge theory to a dual model; this procedure allows to get high precision results.

The structure of the paper goes as follows: in section \ref{effectivestringtheorysect} the effective string scenario for confinement is described; next, in section \ref{u1sect} the basic properties of compact U(1) lattice gauge theory are reviewed. Section \ref{dualitysect} presents the duality transformation mapping the U(1) theory to a model with integer variables, while in section \ref{resultssect} a few details about the algorithm and the input parameters are listed; then we discuss the way to measure the different observables, and present numerical results: the latter are compared with theoretical predictions, and their physical meaning is discussed. Finally, in section \ref{conclusionssect} we summarise the main results of the work, stressing some concluding remarks.

An account on preliminary results of this work has been presented at the XXII International Symposium on Lattice Field Theory (Lattice 2004), Batavia, IL, USA, 21--26 June 2004 \cite{Panero:2004zq}.

\section{Effective string theory}\label{effectivestringtheorysect}

The ground state interquark potential $V(r)$ of a heavy $Q\bar{Q}$ pair in a pure gauge theory (without dynamical matter fields) can be expressed in terms of the two-point correlation function $G(r)$ among Polyakov lines:
\eq
\label{defpotenziale}
V(r)=-\frac{1}{L} \log G(r) = -\frac{1}{L} \log \bra P^\dagger (r) P(0) \ket
\en
where $r$ is the interquark distance, while $L$ is the system size in the (compactified) ``time-like'' direction; for systems at zero temperature, the $L \rightarrow \infty $ limit is taken. 

In the confining regime, the asymptotic behaviour of $V(r)$ at large distances is a linear rise, a fact compatible with the idea that the chromoelectric flux lines get squeezed into a ``tube'' between the two charges, whose energy is expected to be (roughly) proportional to its length. It was therefore conjectured that the long distance properties characterising the heavy $Q\bar{Q}$ system in the confinement phase can be modelled in terms of a vibrating, string-like object joining the two sources; this approach, which nicely fits the intuitive picture of a chromoelectric flux tube squeezed into a thin Abrikosov-Nielsen-Olesen vortex \cite{Abrikosov:1956sx,Nielsen:1973cs}, was quantitatively developed by L\"uscher, M\"unster, Symanzik and Weisz \cite{Luscher:1980fr,Luscher:1980iy,Luscher:1980ac} more than two decades ago.

Obviously, in a realistic model, matter fields would be dynamically created out of the vacuum and induce the ``string breaking'' phenomenon, leading to the flattening of the interquark potential at large distances; however, here we restrict our attention to the case of a pure gauge theory, with no dynamical matter fields.

According to the effective string scenario, the IR behaviour of $V(r)$ should be predicted from the form of the action governing the dynamics of the string-like flux tube between the static sources. In general, the contributions beyond the linearly rising term in $V(r)$ are expected to depend on the functional form of the effective string action, but the coefficient of the $1/r$ contribution (which is known as the ``L\"uscher term'', the leading string correction in the IR regime) can be worked out under the very general assumption that, at large distances, the effective model reduces to a free vibrating string. A first quantisation approach shows that the leading contributions to $V(r)$ in the string picture read \cite{Luscher:1980fr}:
\eq
\label{potenziale}
V(r)=\sigma r + \mu - \frac{\pi (D-2)}{24r} + O\left( \frac{1}{r^2} \right)
\en
where $D$ is the number of spacetime dimensions, $\sigma$ is the string tension, and $\mu$ is a constant. The L\"uscher term in eq. (\ref{potenziale}) must not be confused with a Coulomb term originating from the one-gluon exchange process (which is relevant at short distances); on the contrary, the L\"uscher term is a Casimir effect due to the finiteness of the interquark distance $r$, and it is a consequence of the hypothesis that, at leading order, the transverse string fluctuations are described by a CFT of $(D-2)$ free, massless boson fields. Actually, it was shown that $ - \frac{\pi (D-2)}{24} $ is a universal coefficient that one expects in any model based on a string theory \cite{Luscher:1980ac,Stack:1980zn}.

Whether the effective string model provides a satisfactory picture or not, is still a question under debate. On one hand, the interquark potential $V(r)$ (and related quantities) have been investigated in a large number of lattice studies for different gauge theories since the Eighties --- see \cite{Bali:2000gf,Bali:2004tm} for references --- with results confirming the bosonic string prediction: in particular, the L\"uscher term appears to be a universal feature of the IR regime of confined gauge theories. On the other hand, data from recent, high precision Monte Carlo simulations \cite{Caselle:2002rm,Luscher:2002qv,Caselle:2002ah,Majumdar:2002mr,Caselle:2003rq,Caselle:2003rs,Koma:2003gi,Caselle:2004jq,Caselle:2004er,Majumdar:2004qx,Caselle:2005xy} have shown that higher order corrections to $V(r)$ might be \emph{non-universal}, and --- in general --- do not trivially fit the bosonic string predictions. 

A related aspect, which is still under debate, concerns the excitation spectrum; according to the bosonic string picture, it is expected to be described by a tower of harmonic oscillator excitations with energies:
\eq
\label{spettrooscarm}
E=E_0 + \frac{\pi}{r}n \;\; , \;\;\;\;\;\;\;\; n \in \N
\en
where degeneracies are expected for $n \ge 1$. However, while the L\"uscher term in the ground state potential appears to be confirmed by numerical evidence down to surprisingly short distances, the status of lattice simulation results for the excited states seems to be much more difficult to fit into this simple harmonic oscillator pattern: in some studies \cite{Juge:1999ar,Juge:2002br,Juge:2003sz,Juge:2004xr} it was found that, at distance scales around $0.5$ to $1.0$ fm (where the L\"uscher term in the ground states is observed) the excitation spectrum of the static interquark potential does not agree with eq. (\ref{spettrooscarm}). A similar mismatching between effective string predictions \cite{Ambjorn:1984yu} and numerical results was also observed for the torelon states \cite{Juge:2003vw}.

Another prediction of the effective string models concerns the behaviour of the ``flux tube width'' as a function of the distance between the sources \cite{Luscher:1980iy}: in the confinement phase, the gluon field in the ground state of a heavy $Q\bar{Q}$ pair is essentially expected to be non-vanishing inside a tube-like region between the charges, and to be exponentially decreasing with the distance from the interquark axis. A measure of the (square) typical transverse width of the flux tube can be expressed as:
\eq
\label{squarewidth}
w^2 = \frac{\int d^2 x_{\perp} x_{\perp}^2 \mathcal{E}(x)}{\int d^2 x_{\perp} \mathcal{E}(x)}
\en
evaluated in the mid-plane between the charges, where $x_{\perp}$ is the transverse coordinate, while $\mathcal{E}(x)$ is the chromoelectric field energy density of the $Q\bar{Q} $ ground state with respect to the vacuum:
\eq
\label{chromoelectricfieldenergydensity}
\mathcal{E}(x) \propto \bra Q\bar{Q} | \tr \vec{E}^2(x) | Q\bar{Q} \ket - \bra Q\bar{Q} | Q\bar{Q} \ket \bra 0 | \tr \vec{E}^2(x) | 0 \ket
\en
In the effective string scenario, the flux tube width measures the amplitude of string vibrations, and, in the Gaussian approximation, it is expected to increase logarithmically as a function of the interquark distance:
\eq
\label{squarewidthstringprediction}
w^2 = w^2_0 \ln \left( \frac{r}{\lambda} \right)
\en
(where $\lambda$ is a length scale, and $w^2_0$ is a constant); this behaviour was indeed confirmed by numerical simulations of $\Z_2$ gauge model in $D=3$ \cite{Caselle:1995fh} and U(1) gauge model in $D=4$ \cite{Zach:1997yz}.

Focusing the attention onto the ground state interquark potential, the prediction for $V(r)$ of a $Q\bar{Q}$ pair eq. (\ref{potenziale}) can be refined as it follows. Assuming that the spectrum (including multiplicities) of the pure gauge model is described by the effective string, and neglecting excited state decays, the partition function for this sector of the theory is given by the Polyakov loop two-point correlation function $G(r)$. The latter is written as a string partition function:
\eq
\label{prp0conazeff}
G(r) = \bra P^\dagger (r) P(0) \ket = \int \left[ \mathcal{D} h \right] e^{-S_{\mbox{\tiny{eff}}}} 
\en
where $S_{\mbox{\tiny{eff}}}$ is the effective action for the world sheet spanned by the string, and functional integration is done over world sheet configurations having the Polyakov lines as a fixed boundary. At leading order, one can assume that the degrees of freedom associated with transverse string vibrations are free, massless bosons\footnote{This hypothesis is justified in the so-called ``rough'' regime \cite{rough} of the theory.}, in agreement with the physical content of eq. (\ref{potenziale}).

To get more detailed information about the subleading terms that do not appear in eq. (\ref{potenziale}), one should know the higher-order contributions to the effective string action, which introduce string interactions. 

Choosing $S_{\mbox{\tiny{eff}}}$ to be proportional to the area spanned by the string world sheet one gets the Nambu-Goto action \cite{Nambu:1970,Goto:1971ce,Nambu:1974zg}:
\eq
\label{azionenambugoto}
S_{\mbox{\tiny{eff}}}= \sigma \cdot \int d^2 \xi \sqrt{ \det g_{\alpha \beta} }
\en
This string action suffers from inconsistency problems at a quantum level: rotational invariance is broken once one considers the problem from the quantum point of view, unless the number of spacetime dimensions is $D=26$ \cite{Goddard:1973qh}. 

In the \emph{physical gauge}\footnote{Since eq. (\ref{azionenambugoto}) is invariant with respect to reparametrization and Weyl transformations, a gauge choice is needed to proceed to quantum treatment. In the physical gauge, two degrees of freedom are associated with the ``natural'' coordinates, and the remaining ones describe transverse displacements. We neglect the problems related to Weyl anomaly, since the latter vanishes at large distances.} the integrand appearing in eq. (\ref{azionenambugoto}) can be written as:
\eq
\label{azionenambugoto2}
\left[ 1 + ( \partial_0 h )^2 + ( \partial_1 h )^2  + ( \partial_0 h \times \partial_1 h )^2 \right]^{\frac{1}{2}}
\en
($h$ being the transverse displacement of the world sheet surface). 

Although the resulting theory is non-renormalizable, expanding eq. (\ref{azionenambugoto2}) in a perturbative series, the (regularised) expression for $\bra P^\dagger (r) P(0) \ket$ at leading order reads:
\eq
\label{nambugotoleadingorder}
\bra P^\dagger (r) P(0) \ket = \frac{e^{- \sigma rL - \mu L}}{\left[ \eta \left( i \frac{L}{2r} \right)\right]^{D-2}}
\en
where Dedekind's $\eta$ function:
\eq
\label{dedekind}
\eta ( \tau ) = q^{\frac{1}{24}} \prod_{n=1}^{\infty} (1-q^n) \;\; , \;\;\;\;\;\;\;\; \mbox{with: } \;\;\;\; q = e^{2 \pi i \tau}
\en
encodes the contribution of massless fluctuations. Notice that --- as it is expected --- eq. (\ref{nambugotoleadingorder}) reproduces the L\"uscher term contribution to the interquark potential in the ``low temperature regime'' $\frac{L}{2r} \gg 1$.

The exact spectrum of the Nambu-Goto string can be obtained via formal canonical quantisation \cite{Arvis:1983fp} (see also \cite{Alvarez:1981kc}); the energy levels for a string with fixed ends read: 
\eq
\label{enarvis}
E_n(r)= \sigma r \sqrt{ 1 + \frac{2 \pi}{ \sigma r^2} \left( n - \frac{D-2}{24} \right)  } \;\; ,  \;\;\;\; n \in \N
\en
The string partition function can be written in terms of these energy levels as: 
\eq
\label{partitionfunctionasaseries}
Z = \sum_{n=0}^{\infty} w_n e^{- E_n(r) L}
\en
with some integer weights $w_n$, which account for state multiplicities. The simplest example is the $D=3$ case, where the $w_n$ coefficients are the partitions of $n$:
\eq
\label{partitionfunctionasaseries3d}
Z = \sum_{n=0}^{\infty} P(n) e^{- E_n(r) L}
\en
which has been recently compared with numerical results of $\Z_2$, SU(2) and SU(3) gauge theories \cite{Caselle:2005xy}.

The picture of an effective string can be generalised in different ways: for instance, in \cite{Polchinski:1991ax}, Polchinski and Strominger proposed a non-polynomial action:
\eq
\label{polchinskistromingereffectiveaction}
S_{\mbox{\tiny{eff}}}= \frac{1}{4\pi} \int d \tau^+ d \tau^- \left[ 
\frac{1}{a^2}(\partial_+ X \cdot \partial_- X )  
+ \left( \frac{D-26}{12} \right) \frac{ ( \partial_+^2 X \cdot \partial_- X ) (\partial_+ X \cdot \partial_-^2 X )}{(\partial_+ X \cdot \partial_- X )^2} + O(r^{-3})\right]
\en
(where we followed the notation of the original paper); here, $a$ denotes the typical length scale related to the string tension. In $D=4$, eq. (\ref{polchinskistromingereffectiveaction}) describes an interacting, Poincar\'e-invariant CFT with four bosons and central charge $c=26$; the second term appearing in the integrand on the right hand side of eq. (\ref{polchinskistromingereffectiveaction}) is related to the determinants discussed by Polyakov in \cite{Polyakov:1981rd}, although here they are meant to be built out of the induced metric instead of the intrinsic one. More precisely, the idea underlying the construction of the effective action given by eq. (\ref{polchinskistromingereffectiveaction}) is to allow a generic (but conformally invariant) world-sheet QFT, and to absorb the Jacobian into the action coefficients; the latter can be fixed in such a way that the conformal anomaly is cancelled in any dimension $D$. The non-polynomial terms in eq. (\ref{polchinskistromingereffectiveaction}) are not pathological, provided one considers an expansion about the long-string vacuum\footnote{When the string length is much larger than $a$, the infinite number of terms originating from the expansion of eq. (\ref{polchinskistromingereffectiveaction}) are irrelevant, as they are suppressed by powers of $a$.}. It has been recently pointed out \cite{Drummond:2004yp} that Poincar\'e invariance also constrains the $O(r^{-3})$ term of the string spectrum to be exactly the same that shows up in the Nambu-Goto model.

Eq. (\ref{azionenambugoto}) and eq. (\ref{polchinskistromingereffectiveaction}) can be considered as particular examples for a candidate effective string action; more generally, the effective string action is expected to be written as a series like \cite{Luscher:2002qv,Luscher:2004ib}:
\eq
\label{serieseff}
S_{\mbox{\tiny{eff}}}= \sigma rL + \mu L + S_0 + S_1 + S_2 + \dots , \;\;\;\;\;\; \mbox{with: } \;\;\;\; S_0=\frac{1}{2} \int d^2 \xi (\partial_a h \partial_a h ) 
\en
$S_0$, the leading term (beyond the ``classical'' contributions), describes a conformal model, while the subsequent $S_n$ terms, with $n \ge 1$, have increasing dimension, and introduce string self-interactions. 

In particular, $S_1$ is associated with a ``boundary contribution'' \cite{Luscher:2002qv}:
\eq
\label{s1}
S_1 = \frac{b}{4} \int d \xi_0 \left[ (\partial_1 h)_{\xi_1=0}^2 + (\partial_1 h)_{\xi_1=r}^2  \right]
\en
$b$ being a coupling with length dimension. A perturbative calculation \cite{Caselle:2004jq,Luscher:2004ib} shows that the effect of $S_1$ amounts to a shift $r \rightarrow r-b $ in the contribution induced by the $S_0$ term\footnote{In particular, this holds up to and including $O(b^2)$ terms.}. The role possibly played by this ``boundary contribution'' has been studied in detail in some recent works: on the numerical side, comparison of simulation results for different gauge models $D=3$ and $D=4$ seems to suggest that the $S_1$ contribution is not universal \cite{Koma:2003gi,Juge:2004xr,Caselle:2004jq,Caselle:2004er} --- as opposite to the L\"uscher term induced by $S_0$. On the theoretical side, it has been recently pointed out \cite{Luscher:2004ib} that the $O\left( r^{-2} \right)$ contribution induced by $S_1$ has to be ruled out, if open-closed string duality holds.

The effective string scenario described here refers to a ``meson-like'' $Q\bar{Q}$ system, but it can be generalised to the case of ``baryonic'' structures of three quarks as well \cite{Jahn:2003uz}; however, here we do not address this issue, restricting our attention to predictions for the $Q\bar{Q}$ system that we studied in lattice simulations.

\section{Compact U(1) lattice gauge theory}\label{u1sect}

We study compact U(1) gauge theory in four spacetime dimensions, on a hypercubic, isotropic, Euclidean lattice $\Lambda$ with spacing $a$; periodic boundary conditions are imposed along all directions; here and in the following, $N_\mu$ denotes the number of lattice sites along the $\mu$ direction.

This simple abelian theory has some interesting properties in common with non-abelian models --- remarkably, the existence of a confined phase --- and, historically, it has been the object of many numerical studies since the early days of lattice gauge theory. The analogy with its three-dimensional counterpart, where charge confinement mechanism is driven by condensation of monopole plasma \cite{Polyakov:1975rs,Polyakov:1976fu}, is supported by numerical evidence that the monopole density is non-vanishing in the confined phase (at strong coupling), while it goes to zero in the deconfined regime.

The fundamental degrees of freedom of the model are $U_\mu(x) = \exp \left[ i \theta_\mu(x) \right] \in \mbox{U(1)}$ variables defined on the oriented links of the lattice; their dynamics is governed by the standard Wilson action:
\eq
\label{wilsonaction}
S=\beta \sum_{\mbox{\tiny{pl.}}} \left[ 1 - \cos\left( d \theta \right) \right]
\en
where the summation ranges over the lattice plaquettes, $d$ denotes the discretised exterior derivative, and $\beta=\frac{1}{e^2}$.

At zero temperature, the theory has two different phases: for $\beta$ values below a critical value $\beta_c = 1.0111331(21)$ \cite{Arnold:2002jk}, the system is in the \emph{confined phase}, whereas for $\beta > \beta_c$ it lies in a \emph{Coulomb-like phase}; the transition at $\beta=\beta_c$ is a (weak) first order one. In the present work, we consider values of $\beta$ in the confined phase, close to the deconfinement transition\footnote{Since the deconfining transition is first order, the correlation length scale $\xi=\sigma^{-\frac{1}{2}}$ is not divergent as the critical point is approached, thus, strictly speaking, one cannot expect to \emph{exactly} recover the continuum limit; however, the limit value of the string tension for $\beta \rightarrow \beta_c$ appears to be quite small (approximately of order $10^{-2} a^{-2}$).}.

As it concerns the finite temperature case, in \cite{Vettorazzo:2004cr} the phase diagram of the theory was studied for different values of the number of Euclidean time slices, using the helicity modulus --- see also \cite{Vettorazzo:2003fg} --- as a tool to investigate the phase transitions: the numerical findings show that in the $(\beta,\frac{1}{N_4})$ plane a phase boundary separates the confining phase from a temporal Coulomb phase for any finite value of $N_4$. Moreover, the critical values of the coupling parameter $\beta_c$ at which the system undergoes a finite temperature phase transition, and the corresponding values of the latent heat were measured; the latter turns out to be a quantity which is finite for large values of $N_4$ (a signal of the first-order nature of the phase transition at $T=0$), but it quickly decreases for growing temperatures. In particular, for $N_4 \le 4$, the latent heat appears to vanish, opening the possibility of a second order phase transition at high temperatures.

The interquark potential for a $Q\bar{Q}$ system can be expressed in terms of the expectation value of the correlator among Polyakov lines $P(\vec{x})$ as:
\eq
\label{defpotenziale}
V(r)=-\frac{1}{N_4 a} \log \bra P^\dagger (r,0,0) P(0,0,0) \ket
\en

We also study the behaviour of the longitudinal component of the electric field $E$ in the $| Q\bar{Q} \ket $ state; $E$ is expressed in terms of the field strength $F$, which can be defined as:
\eq
\label{fieldstrengthf}
F=\frac{\sqrt{\beta}}{a^2} \sin \left( d \theta \right)
\en
This definition is consistent with the electric Gauss law.

\section{Duality transformation}\label{dualitysect}

Compact U(1) lattice gauge theory can be analytically mapped to a dual formulation via a group Fourier transform: the partition function and expectation values of the original theory can be rewritten in terms of variables of the dual model.

This technique is inspired by spin duality in statistical systems \cite{Kramers:1941kn,Savit:1979ny}; in its simplest formulation, it can be used for an arbitrary abelian or solvable group $\mathcal{G}$ \cite{Drouffe:1978xz,Monastyrsky:1978kp} in generic spacetime dimension $D$. This is due to the fact that, for abelian groups, the product of irreducible representations still gives an irreducible representation; for solvable groups, the reasoning can be generalised considering a well-suited mapping of group elements down to an abelian group --- see also \cite{Casalbuoni:1979nw}. On the other hand, a generalisation to non-abelian lattice gauge theory was discussed in \cite{Oeckl:2000hs}, but in that case the resulting spin foam model involves non-trivial constraints.

Among the main features of this duality mapping, it is worth stressing its intrinsically non-perturbative nature, and the fact that it is \emph{not} a configuration-to-configuration mapping.

Besides being theoretically interesting, this mapping also proves to be useful in numerical studies, since evaluation of physical observables from measurements of related quantities in the dual model often offers practical advantages. Improvement may occur due to various reasons: the dual model may be defined in terms of variables which are ``simpler'' (or ``simpler to handle'' for a computer) than the original ones; the dynamics of the dual model itself may be ``simpler'' than in the original theory; some observables may be evaluated with a higher precision from direct measurements in the dual model. Indeed, the duality technique has been successfully used in many numerical studies of $\Z_2$ lattice gauge theory \cite{Gliozzi:1994bc,Caselle:1995wn,Caselle:1996ii,Caselle:2002ah,Juge:2004xr,Caselle:2004jq}, and it was already tested for this model in \cite{Zach:1997yz}, where the electric field, magnetic current, and flux tube width were investigated; on the other hand, in \cite{Polley:1990tf,Cox:1998xj,Jersak:1999nv} it was applied to the Villain formulation of the theory.

As a matter of fact, the duality transformation allows one to trade the continuous variables of U(1) model for integer-valued variables, but the major improvement can be achieved in the study of the interquark force as a function of distance. In the original model, the interquark force is expressed in terms of ratios between expectation values of Polyakov loop correlators, that are exponentially decreasing with the distance $r$; these quantities are thus affected by a fast signal-to-noise ratio decay at large distances, and --- in general --- error-reduction techniques are needed to get high precision results. In the dual formulation, instead, the problem is completely overcome, since a lattice estimate of the interquark force for the U(1) gauge model --- which, \emph{in se}, is finite even for very large interquark distances --- can be obtained from direct measurements in the dual system. In this respect, this duality-inspired method can be considered as a possible alternative to other popular error-reduction numerical techniques, like the multi-level algorithm proposed by L\"uscher and Weisz \cite{Luscher:2001up}, which has been used to achieve exponential error reduction in several different gauge models \cite{Luscher:2002qv,Majumdar:2002mr,Caselle:2004er,Kratochvila:2003zj,Gliozzi:2004cs}, including compact U(1) lattice gauge theory in $D=4$ \cite{Majumdar:2003xm,Koma:2003gi}.

The duality transformation for compact U(1) lattice gauge theory in $D=4$ can be summarised as it follows (see also \cite{Zach:1997yz}): the plaquette terms appearing in the partition function:
\eq
\label{partitionfunction}
Z=\int \prod_{x,\mu} d \theta_{\mu}(x) \exp \left\{ -\beta \sum_{\mbox{\tiny{pl.}}} \left[ 1 - \cos \left( d \theta \right)\right] \right\}
\en
are expanded in a group Fourier series:
\eq
\label{partitionfunction2}
Z=\int \prod_{x, \mu} d \theta_{\mu}(x) \prod_{\mbox{\tiny{pl.}}} \sum_{k \in \Z} \exp \left[ i (k, d\theta) \right] e^{-\beta}
I_{|k|} (\beta)
\en
$I_{|k|} (z)$ is a modified Bessel function of the first kind of order $|k|$, and $k$ is an integer-valued 2-form on the lattice, which is associated with plaquettes. Rewriting the product $(k, d\theta)$ as: $(\delta k, \theta)$, and integrating over the $\theta$ link variables introduces a constraint on the $k$'s:
\eq
\label{partitionfunction3}
Z= (2 \pi)^{4N} \prod_{\mbox{\tiny{pl.}}} \sum_{k \in \Z} \left. e^{-\beta} I_{|k|} (\beta) \right|_{\delta k = 0}
\en
($4N$ being the total number of links). Physically, $k$ is associated to the field strength, and the constraint $\delta k = 0$ enforces the vacuum Maxwell equation:
\eq
\label{maxwelleqvuoto}
\delta F= J_e = 0
\en
The constraint $\delta k=0$ is solved introducing the integer-valued 3-form $l$ such that:
\eq
\label{k=deltal}
k=\delta l
\en
and the partition function, written as a sum over configurations (to be denoted as $ \sum_{ \{ l \} }$) of the ``elementary cube'' variables $l$ taking values in $\Z$:
\eq
\label{partitionfunction4}
Z= (2 \pi)^{4N} \sum_{ \{ l \} } \prod_{\mbox{\tiny{pl.}}} \left[ e^{-\beta} I_{|\delta l|} (\beta) \right] 
\en
can be reformulated in a dual lattice description, where the $l$ 3-form defined in the direct lattice corresponds to a $^\star l$ 1-form living on the oriented links of the dual lattice $^\star \Lambda$:
\eq
\label{partitionfunction5}
Z= (2 \pi)^{4N} \sum_{ \{ ^\star l \} } \prod_{\mbox{\tiny{pl.}}} \left[ e^{-\beta} I_{|d ^\star l|} (\beta) \right]
\en
(in this case, the product ranges over the plaquettes of the dual lattice). Eq. (\ref{partitionfunction5}) describes a model whose fundamental degrees of freedom are $^\star l$ link variables interacting through a plaquette-like term:
\eq
\label{partitionfunction6}
Z=\sum_{ \{ ^\star l \} } \exp \left[ - S_{\mbox{\tiny{dual}}}\right]
\en
which is a sum over all the configurations of integer-valued $ ^\star l$ variables defined on the dual lattice links, with:
\eq
\label{sdual}
S_{\mbox{\tiny{dual}}} = \sum _{\mbox{\tiny{pl.}}} \left[ \beta - \log  I_{|d ^\star l|} (\beta) \right] + \mbox{const}
\en
The expression for $S_{\mbox{\tiny{dual}}} $ is thus given in terms of a sum over the plaquettes of the dual lattice, and it shows that
the dual model is again a ``gauge model'', because $S_{\mbox{\tiny{dual}}} $ only depends on $d ^\star l$ (thus it is left invariant under integer-valued local gauge transformations, defined on the $^\star \Lambda$ sites, that shift the incoming $^\star l$'s by the same amount) and, since $I_{n} ( \beta )$ is a decreasing function of the integer $n$ index, for any fixed, positive value of $\beta$, this interaction has ``ferromagnetic nature'', in the sense that the configurations with the largest statistical weight are the ones making $| d ^\star l |$ minimal.

An analogous expression can be derived when external charges are included; in that case, the starting point is the $Q\bar{Q}$ system partition function:
\eq
\label{zqq}
Z_{Q\bar{Q}}=\int \prod_{x,\mu} d \theta_{\mu}(x) \exp \left\{ -\beta \sum_{\mbox{\tiny{pl.}}} \left[ 1 - \cos \left( d \theta \right)\right] \right\} \prod_{\tau=0}^{N_4 - 1} e^{i \theta_4(r,0,0,\tau)} \prod_{t=0}^{N_4 - 1} e^{-i \theta_4(0,0,0,t)}
\en
which is related to the Polyakov loop two-point correlation function:
\eq
\label{prp0zqqfrattoz}
\bra P^\dagger (r) P(0) \ket = \frac{Z_{Q\bar{Q}}}{Z} 
\en
In complete analogy with $Z$, eq. (\ref{zqq}) can be rewritten in the dual formulation. In this case, the constraint $\delta k = 0$ modifies to $\delta k = \pm 1$ for links belonging to the (anti)charge Polyakov loop, and this can be enforced introducing a further 2-form $n$ which describes a world sheet having the Polyakov lines as its boundary:
\eq
\label{worldsheet}
k = \delta l + n
\en
and which is related to the presence of non-vanishing external charges:
\eq
\label{eqmxconcariche}
e \delta n = J
\en
The precise (local) shape of the stack of non-vanishing $n$'s is irrelevant, and the simplest choice corresponds to the minimal surface tiling 
\begin{figure}
\centerline{\includegraphics[height=100mm]{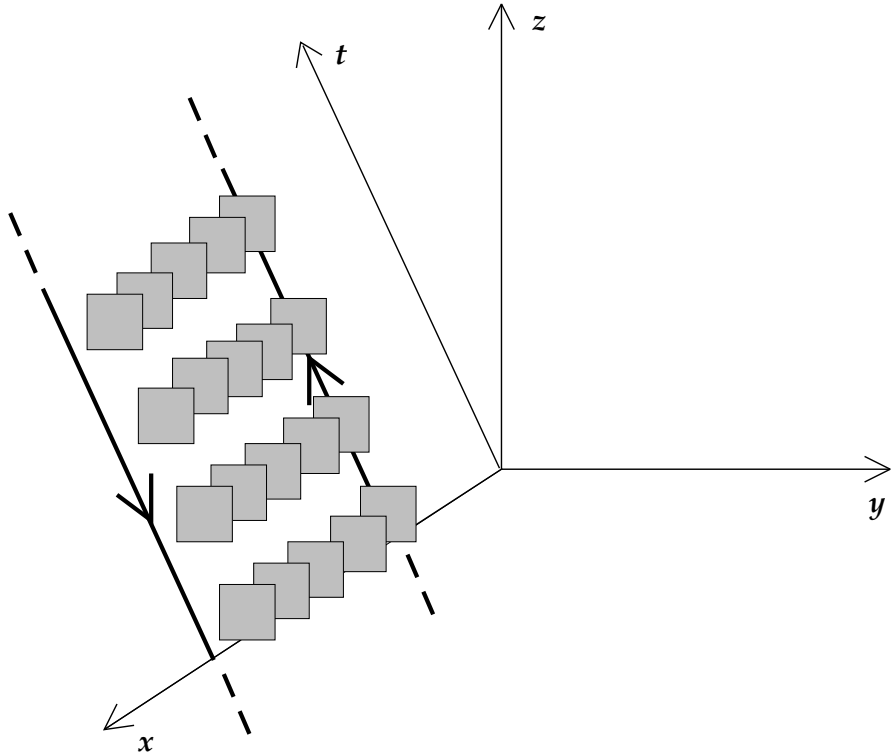}}
\vspace{1cm}
\caption{The $Q\bar{Q}$ pair, described by Polyakov loops in the original model (solid lines), is mapped to a stack of topological defects ($^\star n=+1$) localised onto the shaded plaquettes in the dual lattice. In particular, the shaded plaquettes are dual to an (arbitrary) surface in the original lattice, having the Polyakov lines as its oriented boundary.}
\label{stackfig}
\end{figure}
the sheet between the Polyakov lines --- see also fig.~\ref{stackfig}, showing the $^\star n$ stack of defects as shaded plaquettes in the dual lattice. However, in principle one should keep into account all of the different global (topological) structures for the stack of defects, and include possible $\Z$-twists in the boundary conditions, which create interfaces. The cartoon
\begin{figure}
\centerline{\includegraphics[height=60mm]{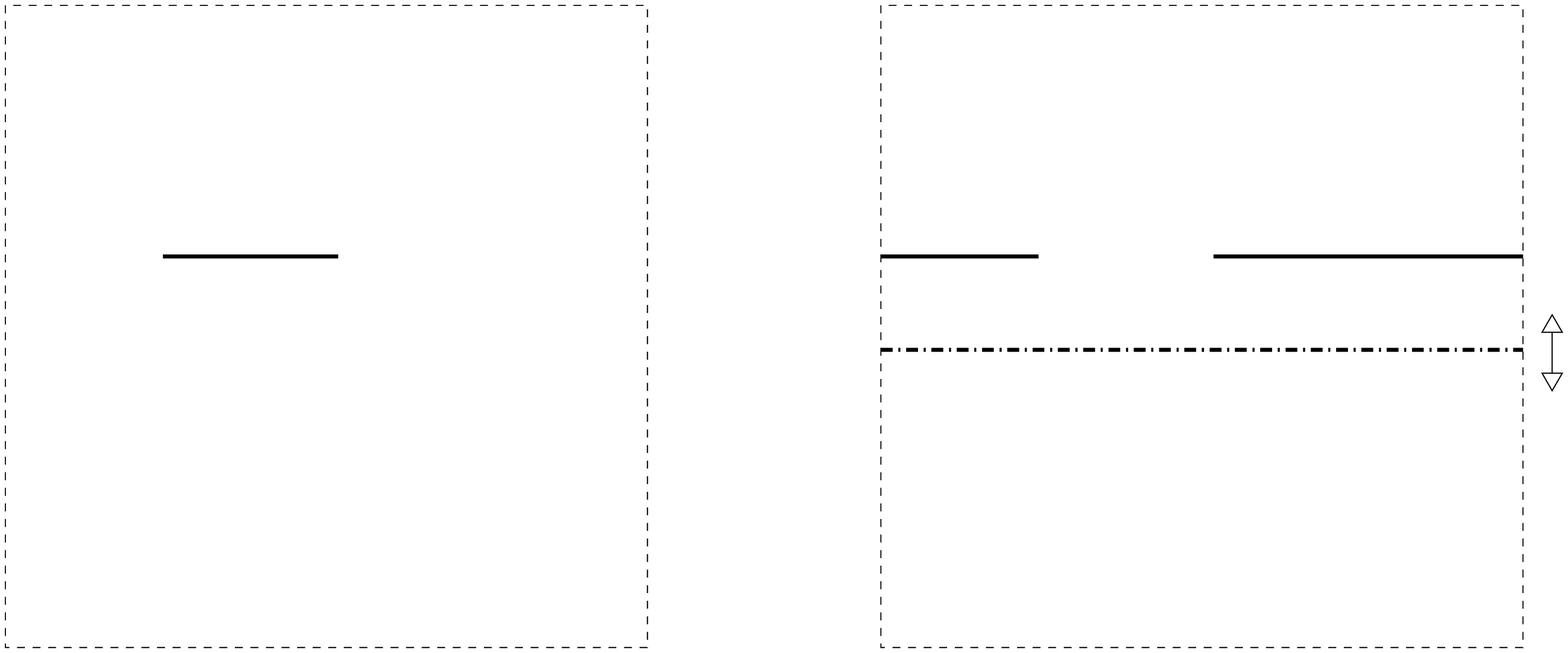}}
\vspace{1cm}
\caption{Schematic, bidimensional, view of two different topological structures for the $^\star n$ stack of defects (solid black lines) in a system with torus-like geometry. On the left side, the defects are located along the shortest line joining the (point-like) sources: this is the analogue of the scheme depicted in fig.~\ref{stackfig}. On the right side, the line of defects joining the sources winds around the system, and a further interface is generated. The latter, by virtue of translational invariance, accounts for an ``entropy gain'', which can compensate for the larger energy cost with respect to the pattern shown on the left side of the figure.}
\label{interfacefig}
\end{figure}
in fig.~\ref{interfacefig} shows a two-dimensional sketch of two different topologies for the $^\star n$ stack: on the right hand side, an interface is created. The larger energy cost associated to this topology can be compensated by the gain in entropy, due to the fact that the interface can be created anywhere. However, non-trivial topologies turn out to be statistically suppressed in the confining regime, for large enough lattice cross-sections, and in our numerical simulations we neglected their effect.

The expression for $Z_{Q\bar{Q}}$ in the dual lattice reads:
\eq
\label{zqqbardual}
Z_{Q\bar{Q}} = \left( 2 \pi \right)^{4N} \sum_{ \{ ^\star l \} } \exp \left\{ - \sum _{\mbox{\tiny{pl.}}} \left[ \beta - \log  I_{|d ^\star l + ^\star n |} (\beta) \right] \right\}
\en
the only difference with respect to eq.~(\ref{partitionfunction6}) being the presence of the $^\star n$ defects which account for the static charges.

Finally, to study the field strength flux through a given plaquette $p$ in the $Q\bar{Q}$ system, one further needs to include the $\sqrt{\beta} \sin \left( d \theta \right)_{p}$ operator in the path integral; the expression for $\bra F_{p} \ket_{Q\bar{Q}}$ can be worked out using the following property of modified Bessel functions of the first kind:
\eq
\label{propertybessel}
\frac{2n}{z} I_n (z) =I_{n-1} (z) - I_{n+1} (z)  
\en
and the result for $ \bra F_{p} \ket_{Q\bar{Q}} $ reads:
\eq
\label{flussodifconqqbar}
\bra F_{p} \ket_{Q\bar{Q}} = \frac{1}{Z_{Q\bar{Q}}} \left( 2 \pi \right)^{4N} \sum_{ \{ ^\star l \} } 
\frac{ ^\star k_{p}}{\sqrt{\beta}} \exp \left\{ - \sum _{\mbox{\tiny{pl.}}} \left[ \beta - \log  I_{|d ^\star l + ^\star n |} (\beta) \right] \right\}
\en

\section{Numerical simulations: settings and results}\label{resultssect}

The algorithm performs Monte Carlo simulations of the dual lattice model, updating the $^\star l$ link variables; the $^\star n$ variables that account for the presence of external charges are kept fixed. We chose the simplest $^\star n$ configuration, like in figure \ref{stackfig}, checking that, for the given values of the parameters, the effect of different topologies could be neglected within the precision of our results.

In the runs, we considered lattice sizes up to $19 \times 31^3$; the $\beta$ coupling --- the parameter of the original theory --- took values from $0.96$ to $1.0107$, in the confined regime and close to the critical value $\beta_c$. The configuration update was done via standard Metropolis method: this choice highlights the advantages arising from the use of the duality transformation.

The autocorrelation time was evaluated using the automatic windowing procedure due to Madras and Sokal~\cite{Madras:1988ei}; then we chose to collect sets of non-autocorrelated data only. The errorbars from data samples have been obtained via the jackknife method.

Detailed results are presented in the following subsections.

\subsection{Interquark potential and force at zero temperature}\label{vrzerotempsubsect}

The interquark potential $V(r)$ and force $F(r)$ can be obtained from the expectation value of the two-point Polyakov loop correlators; we run zero temperature simulations on a hypercubic lattice of size $16^4$, considering different (on-axis) interquark distances. The values of the $\beta$ parameter were chosen in the range between $0.975$ and $1.010$: a window in the confined phase, close to the deconfinement critical point $\beta_c$; correspondingly, the value of the string tension $\sigma$ ranges, approximately, from $ 0.3 a^{-2}$ down to $ 0.06 a^{-2}$, so that the product $ \sigma L^2$ is large enough to damp the statistical weight of configurations with an interface (see discussion in sect.~\ref{dualitysect} and fig.~\ref{interfacefig}). This choice for the parameters also allows a direct comparison with the results obtained with a different algorithm in \cite{Koma:2003gi}.

In particular, it is interesting to note that our algorithm allows to measure easily the ratio between Polyakov loop correlators at distances $r+1$ and $r$: in the dual lattice, such a ratio can be obtained in terms of one column of plaquettes, in the background of a $L \times r$ stack of defects describing the $Q\bar{Q}$ pair at distance $r$. This enables us to overcome the problems due to the fact that $G(r)$ decreases exponentially fast with $r$: although both $G(r+1)$ and $G(r)$ are quantities which become very small already at distances of a few lattice spacings, their ratio goes to a finite limit for large $r$, and it is expected to develop (effective string) corrections for finite values of $r$. In order to achieve a further improvement in the result precision, we actually split the ratio appearing on the right hand side of eq.~(\ref{prp0zqqfrattoz}) into a product of simpler ratios: 
\eq
\label{prodotto}
G(r) = \frac{Z_{Q\bar{Q}}}{Z} = \prod_{i=0}^{L-1} \frac{Z_{\left( L \times r \right)+i+1} }{Z_{\left( L \times r \right)+i}} 
\en
where the generic term $Z_{\left( L \times r \right)+i} $ appearing on the right hand side of eq.~(\ref{prodotto}) describes the partition function of a system where a column of $i$ defects is attached to the original $L \times r$ stack, so that $ Z_{\left( L \times r \right)+i}$ and $ Z_{\left( L \times r \right)+i+1} $ only differ for the presence of a $ ^\star n$ defect on one plaquette. Keeping eq.~(\ref{zqqbardual}) in mind, it is then easy to see that the factors appearing on the right hand side of eq.~(\ref{prodotto}) are nothing but the expectation values of the local quantity:
\eq
\label{placchetta}
\frac{I_{| d^\star l +1 |}(\beta)}{I_{| d^\star l| }(\beta)}
\en
evaluated in a system with a $\left( L \times r \right)+i $ stack of $ ^\star n$ defects. This is the so-called ``snake algorithm'' idea, which was introduced in~\cite{deForcrand:2000fi,Pepe:2001cx}, and that has also been successfully used in~\cite{Caselle:2002ah}. Each of the factors appearing on the right hand side of eq.~(\ref{prodotto}), which are obtained from separate, independent simulations, are affected by small variance.

The first task is to check the presence of the leading-order (LO in the following) correction with respect to the asymptotical linear rise of $V(r)$ which is expected at large distances. In the effective string scenario, in $D=4$ and at zero temperature, eq.~(\ref{nambugotoleadingorder}) gives:
\eq
\label{lostringcorrectiontov}
V(r) = \sigma r + \mu - \frac{\pi}{12 r} + O \left( r^{-2} \right)
\en
Eq.~(\ref{nambugotoleadingorder}) can be fitted to the numerical results for $G(r)$. 
\begin{table}[h]
\begin{center}
\begin{tabular}{|c|c|c|c|c|c|}
\hline
$\beta$ & $r_{\mbox{\tiny{min}}}/a$ & $\sigma a^2$ & $\mu a$ & number of d.o.f. & reduced $\chi^2$ \\
\hline
\hline
0.975 & 1 & 0.31322(95) & 0.4996(33)  & 6 & 0.8 \\
          & 2 & 0.3145(13) & 0.4929(59)  & 5 & 0.7 \\
          & 3 & 0.31208(50) &0.5068(25)  & 4 & 0.6 \\
\hline
0.980 & 1 & 0.28833(79) & 0.5114(29)  & 6 & 1.3 \\
          & 2 & 0.2889(12) & 0.5082(55)  & 5 & 0.9 \\
          & 3 & 0.2869(10) &0.5196(53)  & 4 & 0.8 \\
\hline
0.985 & 1 & 0.2616(15) &0.5189(51)  & 6 & 1.7 \\
          & 2 & 0.2608(24) & 0.523(10)  & 5 & 1.9 \\
          & 3 & 0.25671(97) & 0.5453(50)  & 4 & 0.7 \\
\hline
0.990 & 1 & 0.2277(12) & 0.5273(42)  & 6 & 0.6 \\
          & 2 & 0.2287(18) & 0.5226(83)  & 5 & 0.7 \\
          & 3 & 0.2256(14) & 0.5398(74)  & 4 & 0.3 \\
\hline
0.995 & 1 & 0.20085(62) & 0.5447(22)  & 6 & 0.4 \\
          & 2 & 0.202128(60) & 0.5379(27)  & 5 & 0.2 \\
          & 3 & 0.20184(91) & 0.5396(47)  & 4 & 0.2 \\
\hline
1.000 & 1 & 0.16357(66) & 0.5550(24)  & 6 & 0.4 \\
          & 2 & 0.1634(10) & 0.5561(46)  & 5 & 0.4 \\
          & 3 & 0.16349(88) & 0.5570(49)  & 4 & 0.3 \\
\hline
1.005 & 1 & 0.12252(73) & 0.5718(27)  & 6 & 0.5 \\
          & 2 & 0.1231(11) & 0.5686(49)  & 5 & 0.5 \\
          & 3 & 0.12302(90) & 0.5693(33)  & 4 & 0.8 \\
\hline
1.010 & 1 & 0.06149(87) & 0.5798(32)  & 6 & 0.6 \\
          & 2 & 0.0627(11) & 0.5699(51)  & 5 & 0.7 \\
          & 3 & 0.06291(67) & 0.5734(29)  & 4 & 0.6 \\
\hline
\end{tabular}
\end{center}
\caption{Fit of eq.~(\ref{nambugotoleadingorder}) to the simulation results for the Polyakov loop correlator $G(r)$. Data corresponding to distances below $r_{\mbox{\tiny{min}}}/a$ are discarded.}
\label{fitLOtab}
\end{table}

Results of data analysis for the different values of $\beta$ that were investigated are reported in tab.~\ref{fitLOtab}; since eq.~(\ref{nambugotoleadingorder}) predicts the behaviour at large distances, we progressively discarded the $G(r)$ data corresponding to the smallest $r$ values.
 
The results obtained for the string tension $\sigma$
\begin{figure}
\centerline{\includegraphics[width=0.55\textwidth]{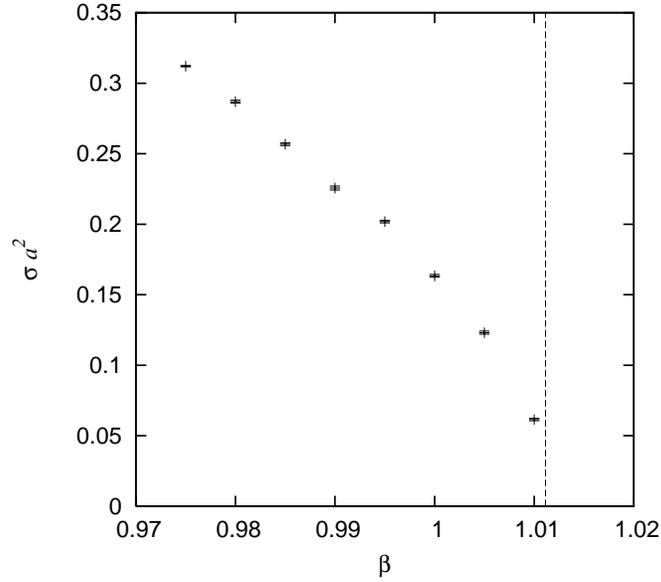}}
\caption{Results for the string tension $\sigma$ (in lattice units), as a function of $\beta$. The dashed line corresponds to the critical value $\beta_c = 1.0111331(21)$ \cite{Arnold:2002jk}.}
\label{stringtensionsfig}
\end{figure}
are plotted as a function of $\beta$ in fig.~\ref{stringtensionsfig}. 

\begin{table}[h]
\begin{center}
\begin{tabular}{|c|c|}
\hline
$\beta$ & $r_{0}/a$  \\
\hline
0.975 & 2.1091(17) \\
0.980 & 2.1997(38) \\
0.985 & 2.3254(44) \\
0.990 & 2.4806(77) \\
0.995 & 2.6225(60) \\
1.000 & 2.9147(89) \\
1.005 & 3.358(15) \\
1.010 & 4.751(34) \\
\hline
\end{tabular}
\end{center}
\caption{Results for Sommer's scale $r_0$ at different values of $\beta$.}
\label{rsommertab}
\end{table}
In tab.~\ref{rsommertab} we also report the results that we obtained for Sommer's scale $r_0$ \cite{Sommer:1993ce}, 
\begin{figure}
\centerline{\includegraphics[width=0.55\textwidth]{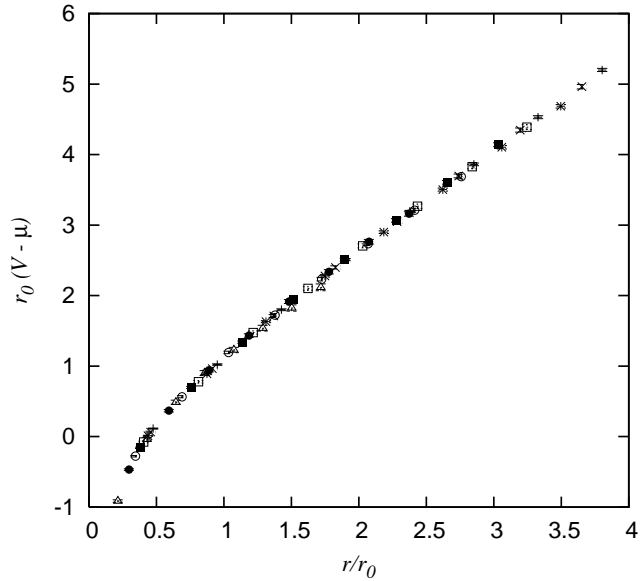}}
\caption{The interquark potential as a function of the distance, in units of Sommer's scale $r_0$. Different symbols refer to different values of $\beta$, in the range from $0.975$ to $1.010$.}
\label{potentialfig}
\end{figure}
which is expected to provide a natural scale to express the numerical results; as an example, in fig.~\ref{potentialfig} the plot for the interquark potential is shown: the good scaling properties of the data obtained from different values of $\beta$ are manifest.

As it concerns the interquark force $F$, evaluated at the mid-point between $r$ and $r+1$, note that it can be approximately estimated in terms of the Polyakov loop correlator ratio $G(r+1)/G(r)$ as:
\eq
\label{approxforce}
F \left( r + \frac{1}{2} \right) \simeq - \frac{1}{L} \log \frac{G(r+1)}{G(r)}
\en

\begin{table}[h]
\begin{center}
\begin{tabular}{|c|c|c||c|c|c||c|c|c|}
\hline
$\beta$ & $r/a$ & $G(r+1)/G(r)$ & $\beta$ & $r/a$ & $G(r+1)/G(r)$ & $\beta$ & $r/a$ & $G(r+1)/G(r)$ \\
\hline
\hline
0.975 & 0 & $1.440(80) \times10^{-4}$ & 0.990 & 0 & $3.66(30) \times 10^{-4}$ & 1.005 & 0 & $9.71(57) \times 10^{-4}$\\ 
          & 1 & $1.002(60) \times10^{-3}$ &           & 1 & $3.93(34) \times10^{-3}$ &            & 1 & $1.90(11) \times 10^{-2}$ \\
          & 2 & $2.70(17) \times10^{-3}$ &             & 2 & $1.069(91)  \times10^{-2}$ &         & 2 & $6.64(39) \times 10^{-2} $\\
          & 3 & $4.49(28) \times10^{-3}$ &             & 3 & $1.74(10)  \times10^{-2}$ &           & 3 & $9.62(57) \times 10^{-2} $\\ 
          & 4 & $5.61(34)  \times10^{-3}$ &            & 4 & $1.99(17)  \times10^{-2}$ &           & 4 & $1.012(60) \times 10^{-1}$\\
          & 5 & $6.13(38)  \times10^{-3}$ &            & 5 & $2.57(22)  \times10^{-2}$ &           & 5 & $1.361(80) \times 10^{-1} $\\
          & 6 & $6.16(38)  \times10^{-3}$ &            & 6 & $2.63(16)  \times10^{-2}$ &           & 6 & $1.317(77) \times 10^{-1}$ \\
          & 7 & $6.15(38)  \times10^{-3}$ &            & 7 & $2.63(16)  \times10^{-2}$ &           & 7 & $1.522(89) \times 10^{-1}$\\
\hline
0.980 & 0 &$1.80(15) \times10^{-4}$  &  0.995  & 0 & $4.20(24) \times 10^{-4}$ & 1.010 & 0 & $2.33(14) \times 10^{-3}$\\ 
          & 1 & $1.39(12) \times10^{-3}$ &             & 1 & $5.54(33) \times 10^{-3}$ &           & 1 & $4.94(29) \times 10^{-2}$\\ 
          & 2 & $4.34(37) \times10^{-3}$ &             & 2 & $2.02(12) \times 10^{-2}$ &           & 2 & $1.642(96) \times 10^{-1}$\\
          & 3 & $6.46(35) \times10^{-3}$ &             & 3 & $2.52(15) \times 10^{-2}$ &           & 3 & $2.42(14) \times 10^{-1}$\\
          & 4 & $8.68(75) \times10^{-3}$ &             & 4 & $3.29(20) \times 10^{-2}$ &           & 4 & $3.24(19) \times 10^{-1}$\\
          & 5 & $8.83(76) \times10^{-3}$ &             & 5 & $3.40(20) \times 10^{-2}$ &           & 5 & $3.46(20) \times 10^{-1}$\\
          & 6 & $8.63(74) \times10^{-3}$ &             & 6 & $3.73(22)  \times10^{-2}$ &           & 6 & $3.73(21) \times 10^{-1}$\\
          & 7 & $1.110(96) \times10^{-2}$ &           & 7 & $3.96(24) \times 10^{-2}$ &           & 7 & $3.66(21) \times 10^{-1}$\\
\hline
0.985 & 0 & $2.53(15) \times10^{-4}$ & 1.000   & 0 & $6.73(40) \times 10^{-4}$ & \multicolumn{3}{|c|}{ } \\
          & 1 & $2.15(13) \times10^{-3}$ &             & 1 & $9.58(58) \times 10^{-3}$ & \multicolumn{3}{|c|}{ } \\
          & 2 & $5.65(34) \times10^{-3}$ &             & 2 & $3.10(19) \times 10^{-2}$& \multicolumn{3}{|c|}{ } \\
          & 3 & $1.074(65)  \times10^{-2}$ &          & 3 & $5.42(32) \times 10^{-2}$& \multicolumn{3}{|c|}{ } \\
          & 4 & $1.298(79)   \times10^{-2}$ &         & 4 & $6.01(36) \times 10^{-2}$ & \multicolumn{3}{|c|}{ } \\
          & 5 & $1.485(90)   \times10^{-2}$ &         & 5 & $6.21(37) \times 10^{-2}$& \multicolumn{3}{|c|}{ } \\
          & 6 & $1.490(90)  \times10^{-2}$ &          & 6 & $7.25(43) \times 10^{-2}$ & \multicolumn{3}{|c|}{ } \\
          & 7 & $1.72(10) \times10^{-2}$ &             & 7 & $7.29(43) \times 10^{-2}$& \multicolumn{3}{|c|}{ } \\
\hline
\end{tabular}
\end{center}
\caption{Ratios of Polyakov loop correlators at different distances and values of $\beta$. These results are obtained from zero temperature simulations on $16^4$ lattices.}
\label{constanterrorstab}
\end{table}

As we already remarked, the results for $G(r+1)/G(r)$ that can be obtained by means of our algorithm at larger and larger distances $r$ are not affected by the typical (exponential) signal-to-noise ratio decay, which is observed for Monte Carlo simulations in the direct model; this is a major advantage that comes from simulations in the dual setting, as it was also pointed out in the study of the $\Z_2$ gauge theory in $D=3$ in \cite{Caselle:2002ah}. 

Tab.~\ref{constanterrorstab} displays our results for $G(r+1)/G(r)$ at different values of $r$ and $\beta$. Notice that, for each set, the uncertainty associated to the data at large values of $r$ appears to be fairly constant, or growing slowly; since the number of measurements to obtain these results was the same for all distances in each set, this improvement in precision with respect to the standard approaches can be fully ascribed to the properties of the algorithm.

According to eq.~(\ref{nambugotoleadingorder}), the LO, large-distance behaviour which is expected for the force reads:
\eq
\label{LOforce}
F \left( r \right) \simeq \sigma + \frac{\pi}{12 r^2}
\en
Expressing our results in terms of Sommer's scale $r_0$, it is possible to make a comparison of the whole set of data obtained from the simulations at values of $\beta$ ranging from 0.975 to 1.005 with the LO prediction obtained from the effective string picture. 
\begin{figure}
\centerline{\includegraphics[width=0.55\textwidth]{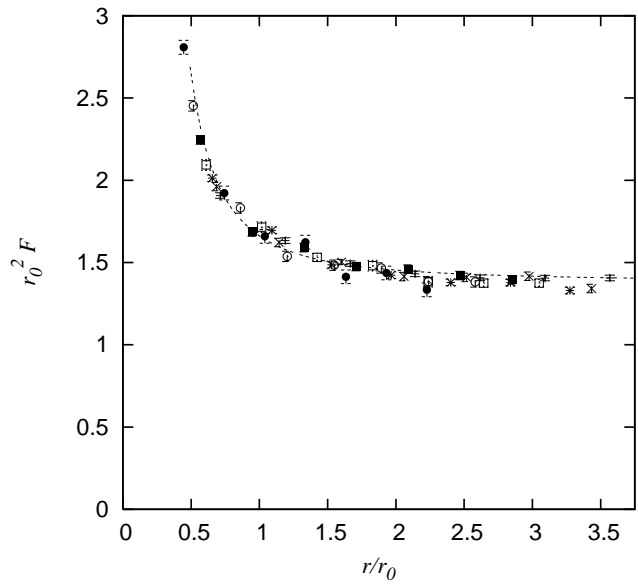}}
\caption{The approximate interquark force $F$ --- obtained from finite (one-lattice-spacing) differences of the potential --- as a function of the (mid-point) distance. All data, obtained from simulations at different values of $\beta$ in the range from $0.975$ to $1.005$, are rescaled in terms of Sommer's scale $r_0$ (which was obtained from separate sets of simulations). The dashed line is the theoretical expectation for the interquark force, as it is predicted by the effective string picture at LO.}
\label{forcefig}
\end{figure}
The corresponding plot is shown in fig.~\ref{forcefig}. Note that, at this level, the dashed curve does not depend on any free parameter, yet it fits quite well with the numerical data, despite the approximations that we discussed. The matching between LO effective string prediction and numerical results appears to hold down to short distances --- a fact in agreement with similar conclusions that have been obtained in other studies of different gauge theories \cite{Necco:2001xg,Luscher:2002qv}.

\subsection{Finite temperature setting}\label{finitetempsubsect}

The analysis of subsection~\ref{vrzerotempsubsect} can be extended to the finite temperature case. There are many physical motivations for addressing the study of gauge theories at finite temperature, and, in particular, it is remarkably interesting to see what happens as one considers temperatures which are closer and closer to the \emph{finite temperature deconfinement transition}: in this subsection, we present the results for the Polyakov loop correlator ratios at finite temperatures, up to about $3/4$ of the critical deconfinement temperature.

The interquark potential $V(r)$ can still be expressed by eq.~(\ref{defpotenziale}), and the LO bosonic string prediction for the Polyakov loop two-point correlation function is given by eq.~(\ref{nambugotoleadingorder}), but an important difference with respect to the zero temperature setting is that, depending on the values of $r$ and $L$, at finite temperatures it is possible to distinguish between two different regimes \cite{Caselle:2002rm,Caselle:2002ah,Caselle:2004er}:
\begin{enumerate}
\item $\frac{L}{2r} > 1$, that we shall call the \emph{low temperature regime};
\item $\frac{L}{2r} < 1$, to be denoted as the \emph{high temperature regime}.
\end{enumerate}
The functional form of the leading term due to string effects in the $\frac{L}{2r} \gg 1$ limit of the low temperature regime obviously reproduces the ``L\"uscher term'' correction to $V(r)$ appearing in the zero temperature case; on the other hand, in order to study the opposite $\frac{2r}{L} \gg 1$ limit of eq.~(\ref{nambugotoleadingorder}), we make use of the following property of Dedekind's function:
\eq
\label{modulareta}
\eta \left( - \frac{1}{\tau} \right) = \sqrt{-i \tau} \; \eta ( \tau)
\en
Accordingly, the LO string prediction for $V(r)$ in the high temperature regime reads:
\eq
\label{vrLOhitemp}
V(r) = \left( \sigma - \frac{\pi}{3 L^2} \right) r + \mu + \frac{1}{L} \log \frac{2r}{L} + \frac{2}{L} \sum_{n=1}^{\infty} \log \left( 1 - e^{- \frac{4 \pi r}{L} n} \right)
\en
from which one can see that the string fluctuations act to renormalize the string tension, through the $- \frac{\pi }{3 L^2} $ term. 

In the following of this subsection, we shall focus our attention on the $2r > L$ case: a regime where numerical studies for the $\Z_2$ gauge model in $D=3$ \cite{Caselle:2002ah} have shown that the finite temperature effects appear to be consistent with the predictions of the Nambu-Goto effective string action truncated at NLO.

In~\cite{Vettorazzo:2004cr}, the values of the critical coupling $\beta_c$ for compact U(1) lattice gauge theory in $D=4$ have been obtained, for different values of the inverse temperature $L$. In particular, here we consider the $L=8$ and $L=6$ cases, for which the values of $\beta$ at which the finite temperature deconfinement transition occurs are, respectively, $\beta_c=1.0107(1)$ and $\beta_c=1.009449(1)$; for these values of the coupling parameter, we run numerical simulations at different temperatures, approaching the deconfinement temperature from below. The lattice size along the remaining directions $N_s$ ($s=1,2,3$) is large enough, so that neglecting non-trivial topologies for the stack of $^\star n$ defects is still a justified approximation.

\begin{figure}
\centerline{\includegraphics[width=0.55\textwidth]{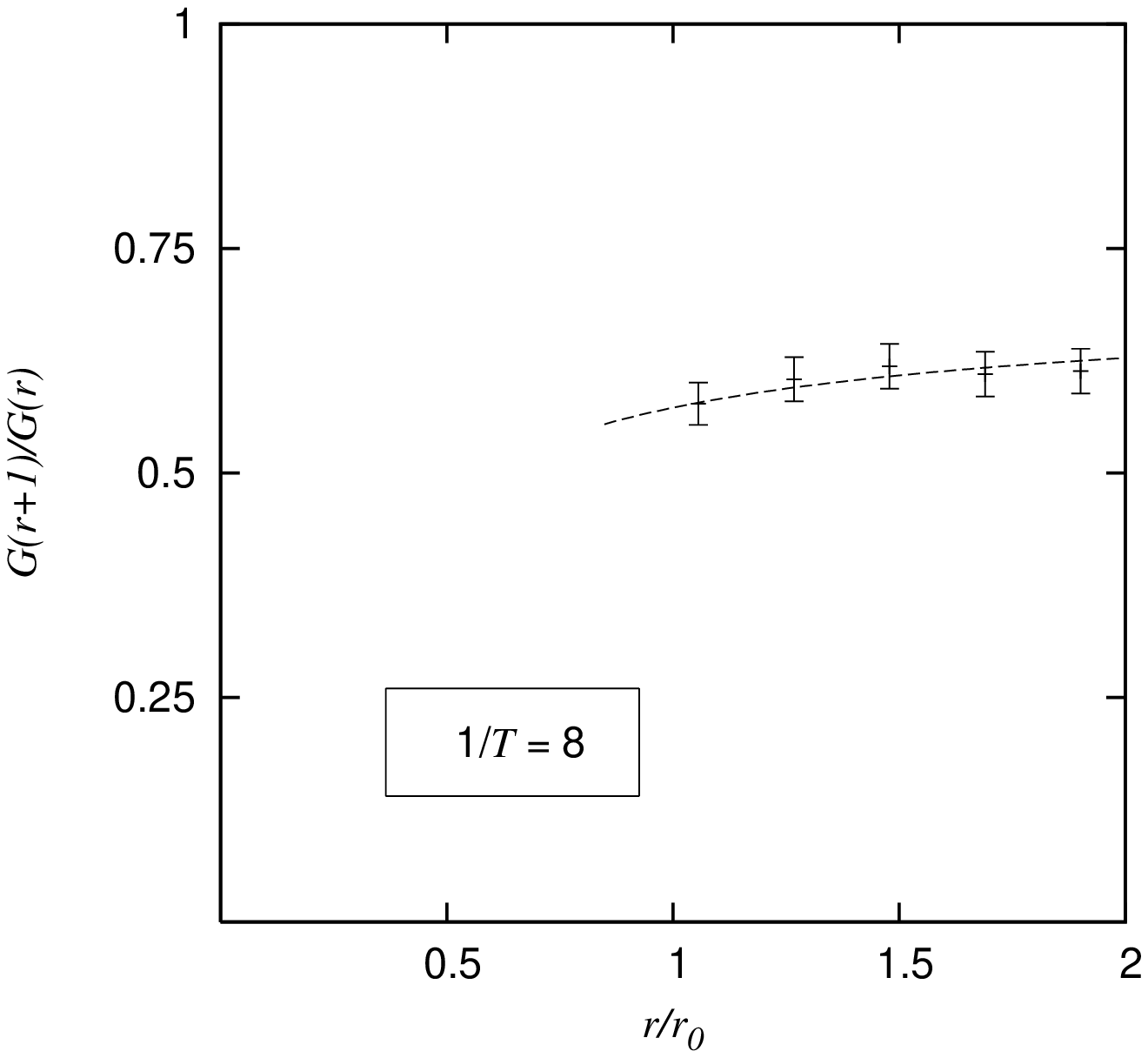}\includegraphics[width=0.55\textwidth]{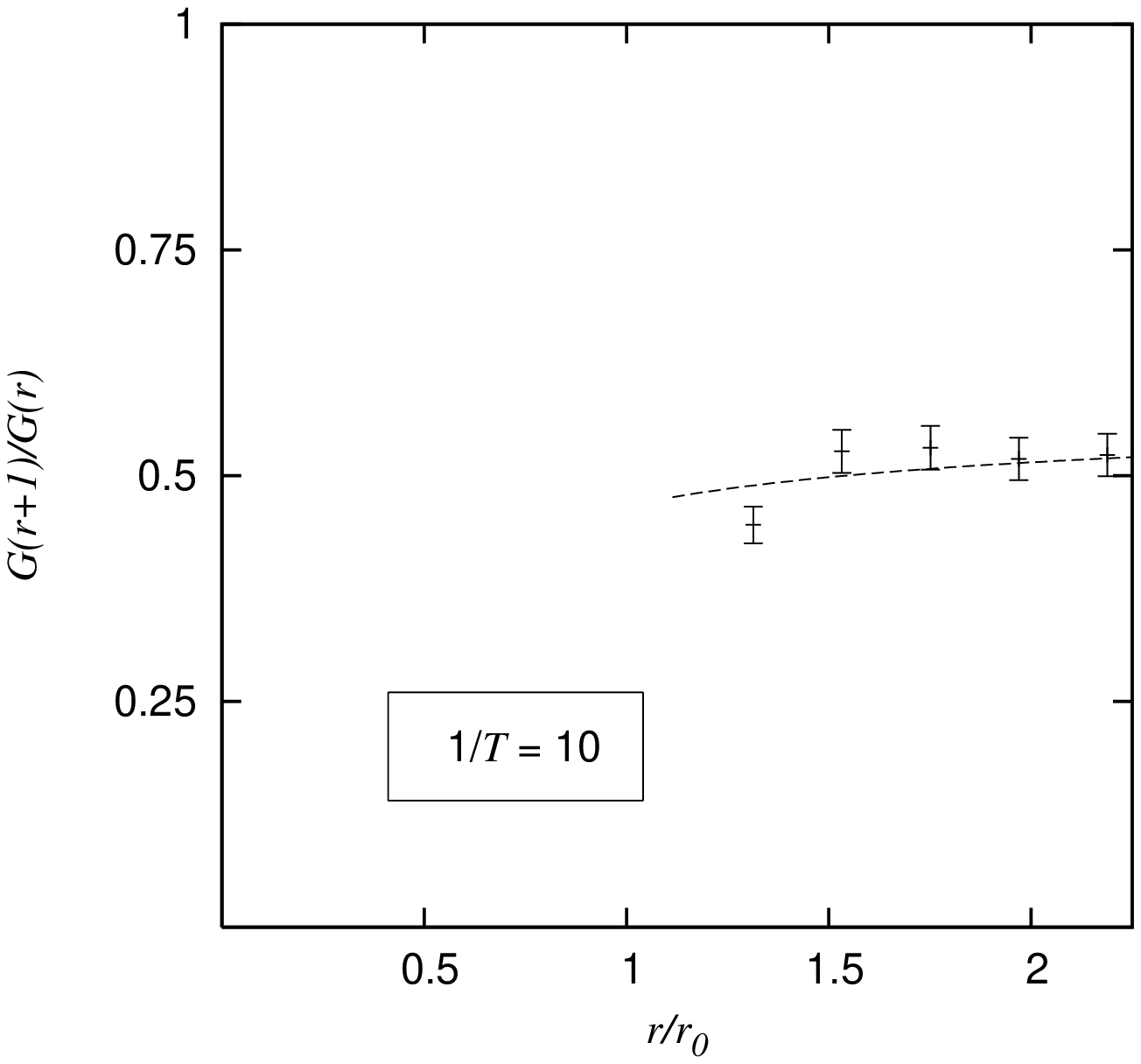}}
\centerline{\includegraphics[width=0.55\textwidth]{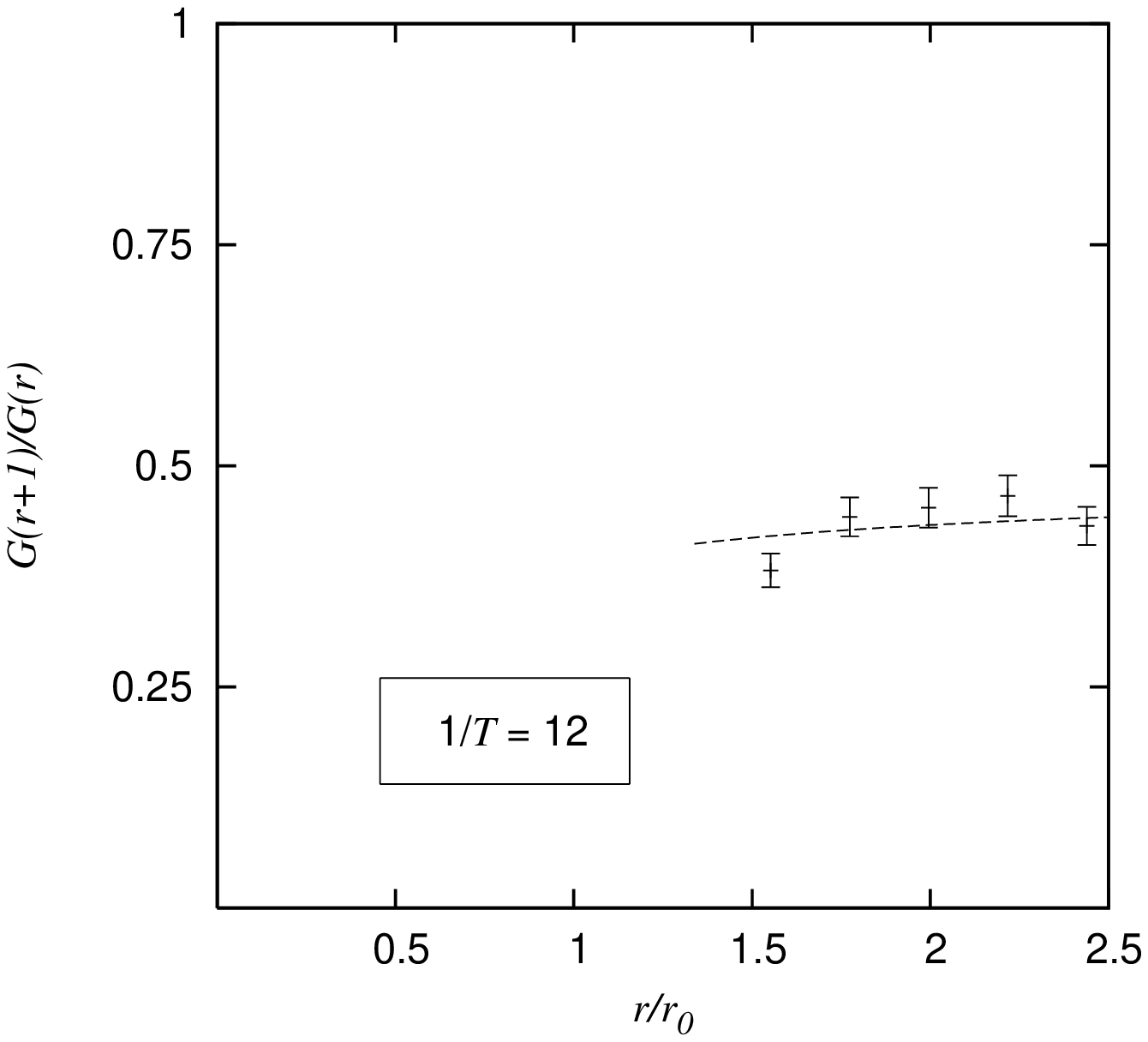}\includegraphics[width=0.55\textwidth]{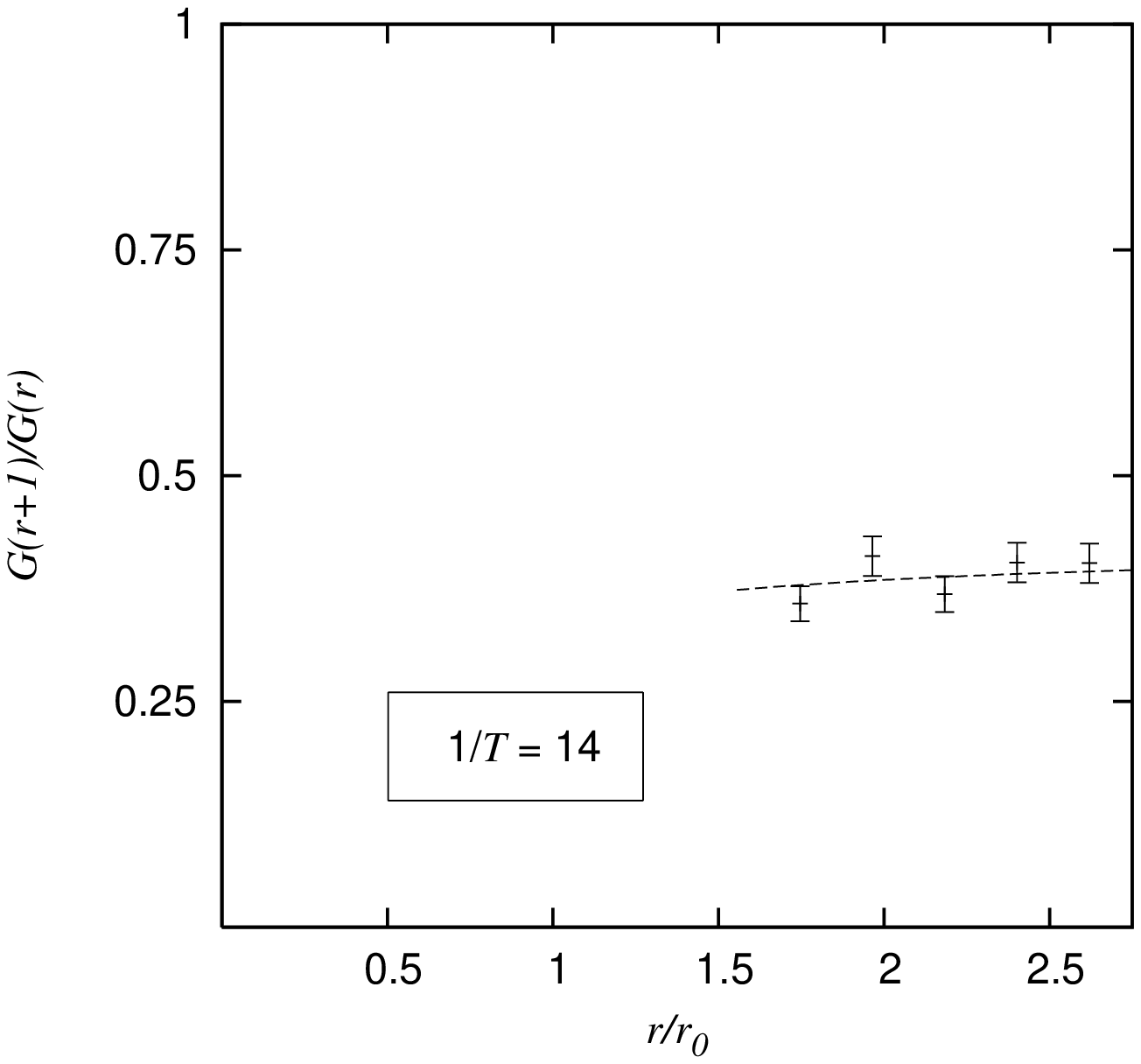}}
\caption{Fit of the theoretical expectation derived from eq.~(\ref{nambugotoleadingorder}) to the numerical results for Polyakov loop correlator ratios, as a function of the interquark distance (expressed in terms of Sommer's scale) at $\beta=1.009449$; the corresponding critical deconfinement inverse temperature is $1/T_c=6$~\cite{Vettorazzo:2004cr}.}
\label{primobetaratiosfig}
\end{figure}

\begin{figure}
\centerline{\includegraphics[width=0.55\textwidth]{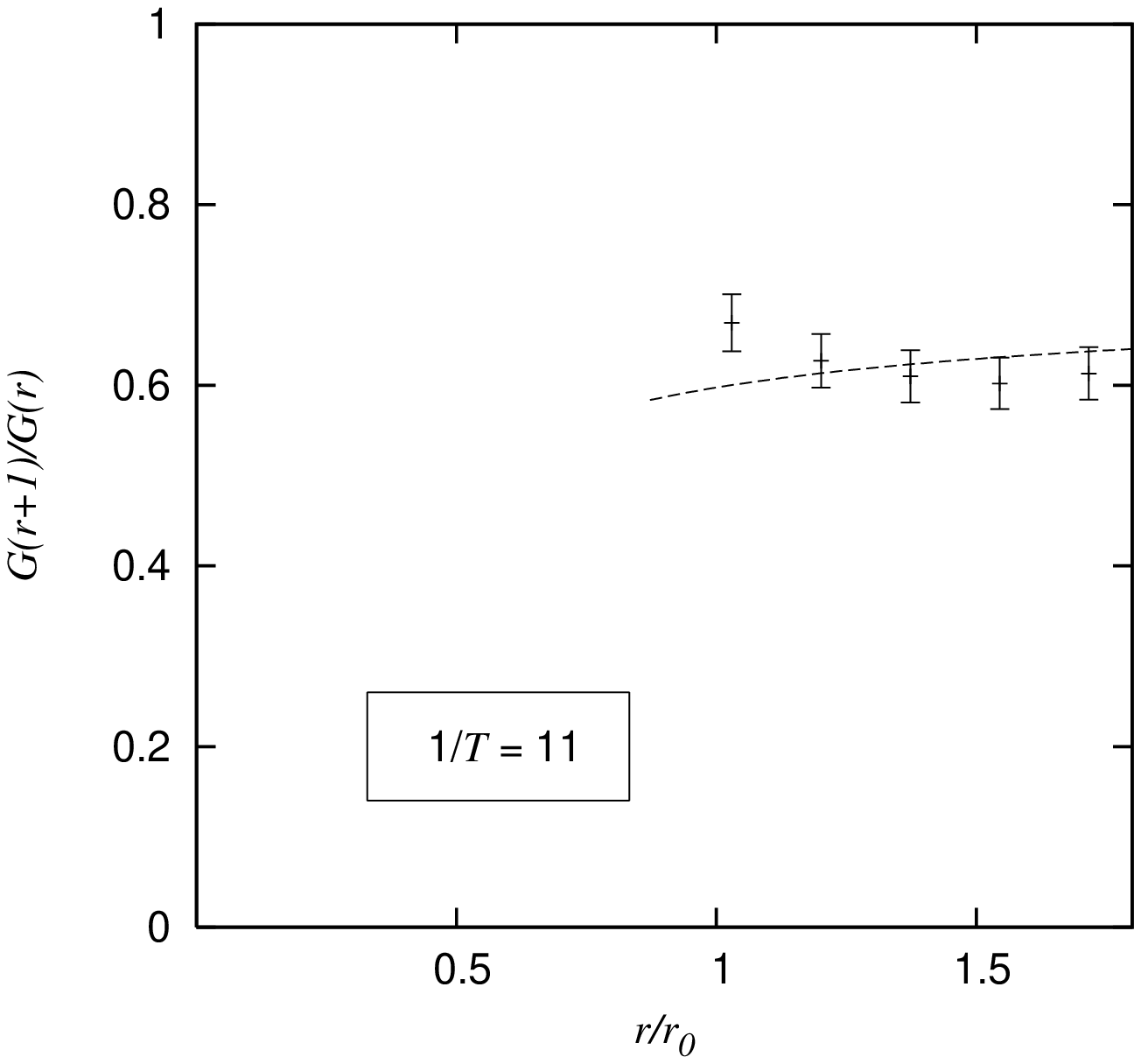}\includegraphics[width=0.55\textwidth]{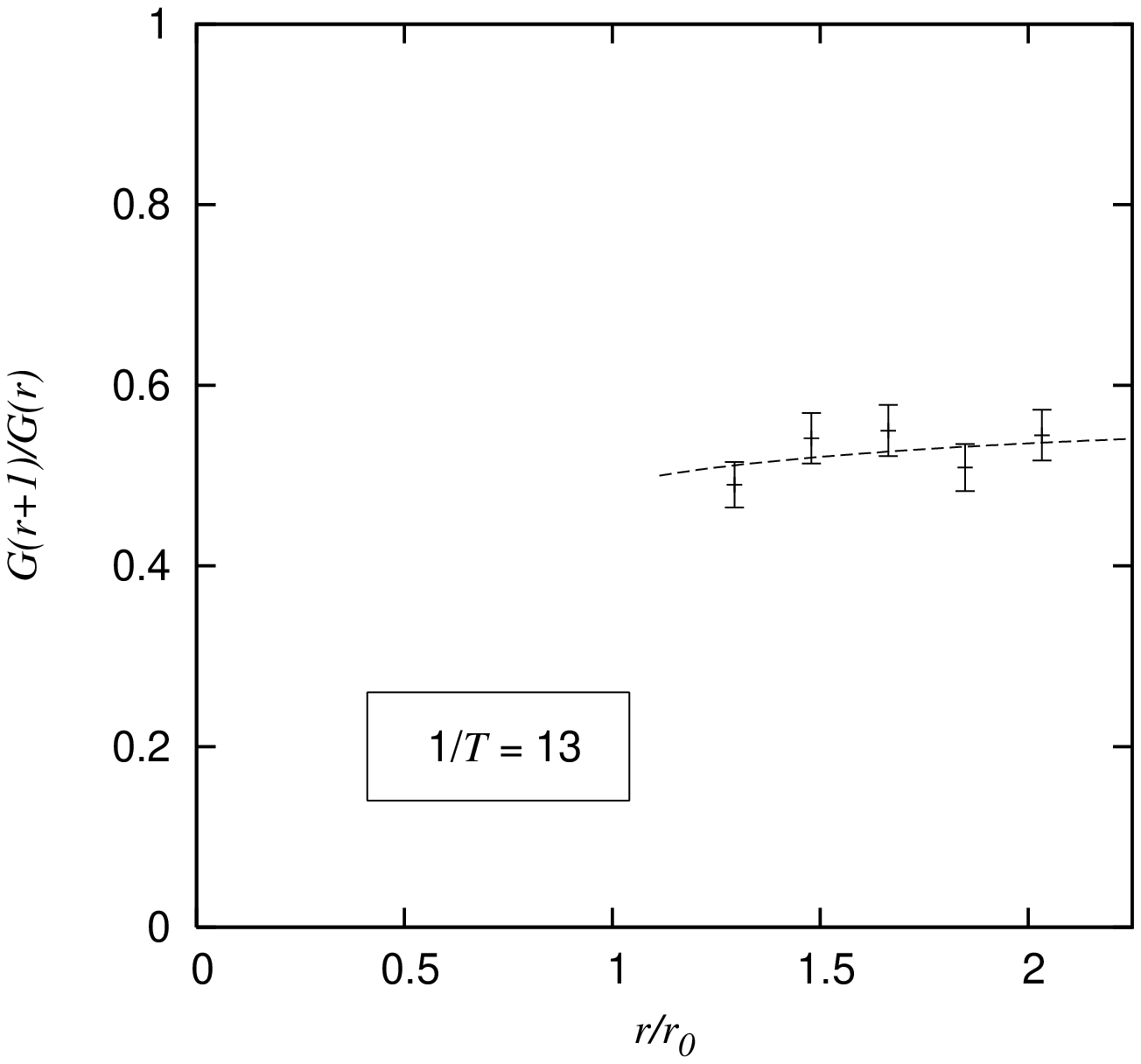}}
\centerline{\includegraphics[width=0.55\textwidth]{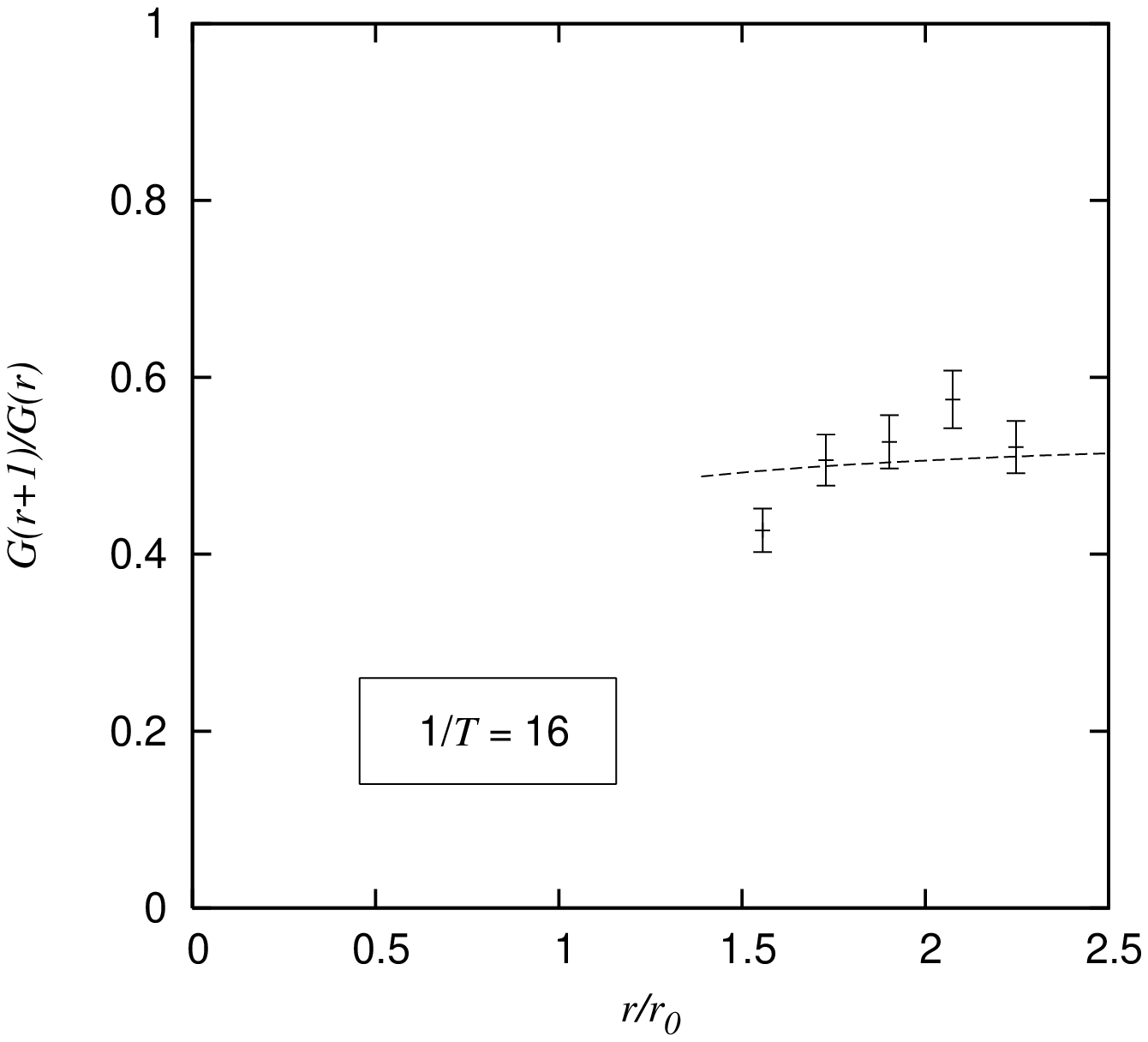}\includegraphics[width=0.55\textwidth]{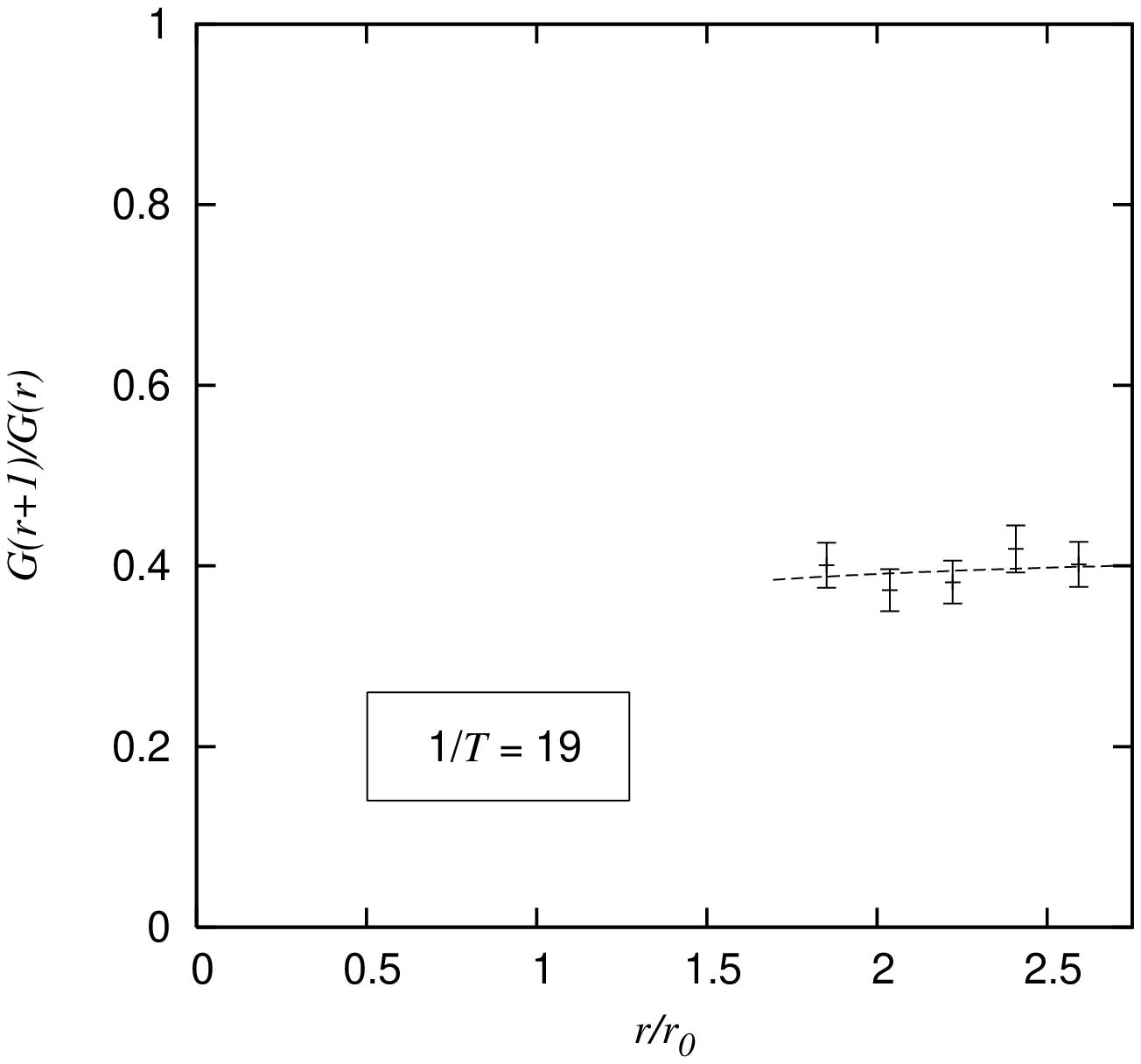}}
\caption{Same as in previous figure, but for the data obtained at $\beta=1.0107$, corresponding to the critical deconfinement inverse temperature $1/T_c=8$~\cite{Vettorazzo:2004cr}.}
\label{secondobetaratiosfig}
\end{figure}

We measured the Polyakov loop correlator ratios $G(r+1)/G(r)$ with the same method discussed in subsection~\ref{vrzerotempsubsect}; information about the parameters in these finite temperature simulations is summarised in tab.~\ref{finitetempinfotab}.
\begin{table}[h]
\begin{center}
\begin{tabular}{|c|c|c|c|c|c||c|c|}
\hline
$\beta$ & $1/T_c$ & $1/T$ & $T/T_c$ & $N_s$ & $r/a$ range & $\sigma a^2$ & reduced $\chi^2$\\
\hline
\hline
1.009449 &  6 &   8 & 0.75 & 21 & 5--9 & 0.06196(91) & 0.16 \\
                    & & 10 & 0.60 & 23 & 6--10 & 0.0665(27) & 1.73 \\
                    & & 12 & 0.50 & 25 & 7--11 & 0.0683(25) & 1.76 \\
                    & & 14 & 0.43 & 27 & 8--12 & 0.0662(17) & 1.02 \\
\hline
1.0107 & 8 & 11 & 0.73 & 23 & 6--10 & 0.0409(25) & 1.74 \\
               & & 13 & 0.62 & 25 & 7--11 & 0.0474(15) & 0.71 \\
               & & 16 & 0.50 & 29 & 9--13 & 0.0415(28) & 3.15 \\
               & & 19 & 0.42 & 31 & 10--14 & 0.0476(10) & 0.47 \\
\hline
\end{tabular}
\end{center}
\caption{Parameter information for the finite temperature runs, together with the  results obtained from the fit of the behaviour predicted by eq.~(\ref{nambugotoleadingorder}) to the $G(r+1)/G(r)$ ratios.}
\label{finitetempinfotab}
\end{table}
The corresponding plots are shown in fig.~\ref{primobetaratiosfig} and in fig.~\ref{secondobetaratiosfig}.

The results displayed in tab.~\ref{finitetempinfotab} are essentially in agreement with the behaviour predicted by eq.~(\ref{nambugotoleadingorder}) for the two-point Polyakov loop correlation function, and the values obtained for the string tension have relative errors between $1\%$ and $3\%$ (with the exception of a unique data sample).

However, data analysis also shows that the precision of these measurements at finite temperature does not allow to give a conclusive answer about the fine effects predicted assuming a string-like behaviour.

\subsection{Electric flux profile}\label{electricfluxprofilesubsect}

In this subsection we discuss the results for the electric field $E$ in the $Q\bar{Q}$ system. With no loss of generality, the $Q\bar{Q}$ axis can be chosen to lie along the $x$ axis of the lattice; we study the electric field longitudinal component $E_x$ in the $(y,z)$ mid-plane between the charges. 

As opposite to the Coulomb-like regime, where $E_x$ is expected to fall off like a power of $\rho=\sqrt{y^2+z^2}$, in the confined phase the dual superconductor picture predicts an exponentially-driven decay which, at large enough distance from the $Q\bar{Q}$ axis, reads \cite{Koma:2003gi}:
\eq
\label{exdualsuperconductor}
E_x \left( \rho \right)= m_B^2 K_0 \left( m_B \rho \right)
\qe
where $m_B$ is the mass of the dual gauge boson. Eq. (\ref{exdualsuperconductor}) is derived in the approximation of an infinitely long flux tube joining the sources, neglecting finite size effects and string fluctuations.

In these simulations, the interquark distance $r$ was kept fixed ($r=3a$), and different values of $\beta$ were chosen, in the range from $0.96$ to $1.01$; different lattice sizes $8^4$ and $16^4$ were also considered, in order to check the influence of possible finite lattice size effects. Information about the parameters is summarised in table~\ref{electrictab}.

\begin{table}[h]
\begin{center}
\begin{tabular}{|c|c|}
\hline
lattice size & $8^4$, $16^4$ \\
\hline
$\beta$ & $0.96$, $0.97$, $0.98$, $0.99$, $1.00$, $1.01$ \\
\hline
interquark distance $r$ & $3a$ \\
\hline
number of measurements &  20~000 \\
\hline
\end{tabular}
\end{center}
\caption{Information about the simulation parameters in the study of the electric flux profile.}
\label{electrictab}
\end{table}

Our data are consistent with other results published in literature; as an example, 
\begin{figure}
\centerline{\includegraphics[width=0.55\textwidth]{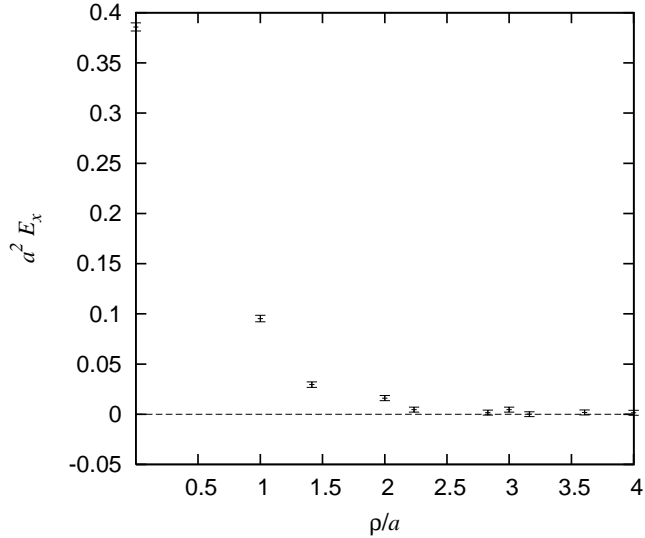}}
\caption{Electric flux through plaquettes in the mid-plane between the charges, as a function of the distance $\rho$ from the $Q\bar{Q}$ axis, at $\beta=0.96$, on a $8^4$ lattice.}
\label{electricfluxfigbeta096lattice8}
\end{figure}
in fig.~\ref{electricfluxfigbeta096lattice8} we show the results that we obtained on a $8^4$ lattice at $\beta=0.96$, which can be directly compared with the ones displayed in the top right corner plot of fig.~2 in \cite{Zach:1995ni}, that were obtained for the same values of the parameters, but from simulations in the direct model.

The data obtained from simulations on a $16^4$ lattice show no relevant discrepancy with respect to the $8^4$ lattice results for the values of $\beta$ that we studied, except for the last one ($\beta=1.01$), which is very close to $\beta_c$: this comes as no surprise, since, in that case, the typical length scale given by the square root of the inverse string tension is of the order of some lattice units (roughly: $\frac{1}{\sqrt{\sigma}} \simeq 4 a$), and the $8^4$ lattice is expected to be affected by some relevant finite size effects. Comparison of results obtained from the $8^4$ and $16^4$ lattices at $\beta=1.01$
\begin{figure}
\centerline{\includegraphics[width=0.55\textwidth]{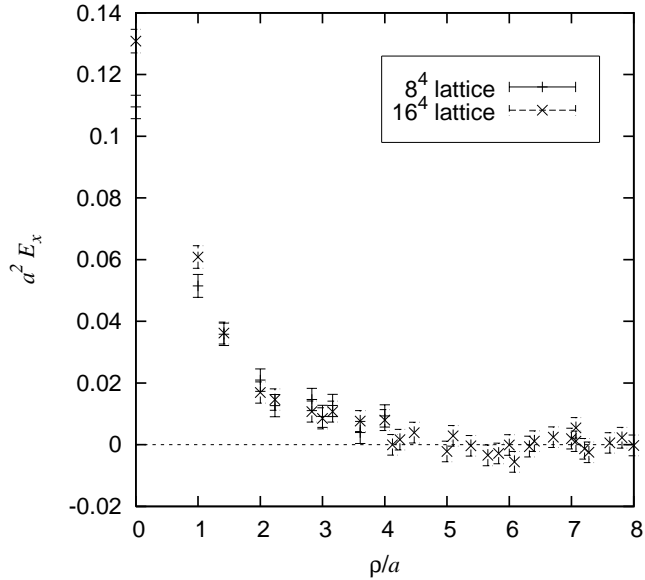}}
\caption{Comparison of results for the longitudinal component of the electric field obtained from simulations on a $8^4$ and on a $16^4$ lattice, at $\beta=1.01$. For the $8^4$ lattice, results are shown up to $\rho/a=4$ only.}\label{electricflux101fig}
\end{figure}
is shown in fig.~\ref{electricflux101fig}.

Testing the prediction of eq.~(\ref{exdualsuperconductor}) is not a trivial task, since it is expected to describe the asymptotic behaviour of $E_x$ at large distance from the interquark axis only, where the signal is decaying exponentially fast with $\rho$; in order to explore large enough distances, we focused our attention onto the data obtained from lattices of size $16^4$. The results obtained from simulations at different values of coupling (ranging from $\beta=0.96$ to $\beta=1.00$) 
\begin{figure}
\centerline{\includegraphics[width=0.55\textwidth]{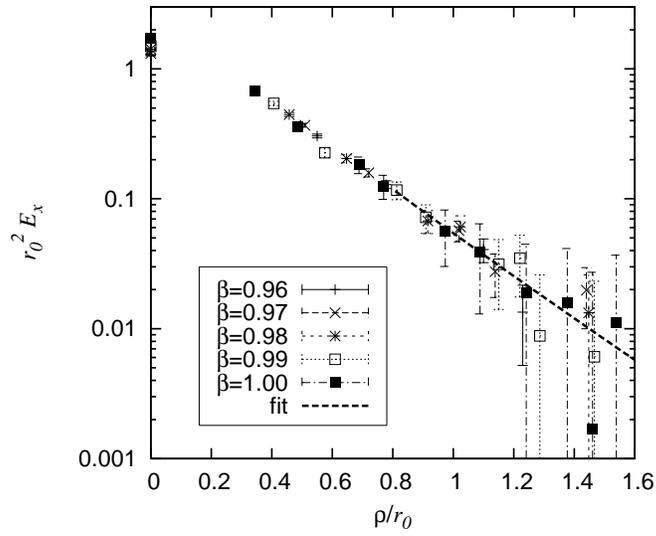}}
\caption{Behaviour of the longitudinal component of the electric field (displayed in a logarithmic scale) as a function of the distance from the interquark axis, for different values of $\beta$ in the confined regime; the results are scaled according to the corresponding values of Sommer's scale $r_0$ (which was obtained from independent measurements). The dashed line is the result of the fit to eq.~(\ref{exdualsuperconductor}), discarding the data which correspond to $\rho/r_0<0.8$.}
\label{newfitfig}
\end{figure}
have a very good scaling behaviour, once they are expressed in terms of Sommer's scale $r_0$, as fig.~\ref{newfitfig} shows: in this semi-logarithmic plot, the data appear to collapse onto a straight line, in agreement with the exponentially decaying asymptotic behaviour predicted by eq.~(\ref{exdualsuperconductor}). As a side remark, please note that fig.~\ref{newfitfig} also shows that the data enjoy good rotational invariance, since results corresponding to various off-axis directions display a behaviour which is fully compatible (within the errorbars) with those measured on the $y$ or $z$ axis. For the values of $\beta$ that are considered here, $r_0$ roughly ranges from $1.8a$ (corresponding to $\beta=0.96$) to $2.9a$ (for $\beta=1.00$, in agreement with \cite{Koma:2003gi}). These results have been  used to extract the dual gauge boson mass $m_B$ from the fit to the behaviour predicted by eq.~(\ref{exdualsuperconductor}); since the latter is expected to describe the asymptotic, large-distance, behaviour only, we discarded 
\begin{table}[h]
\begin{center}
\begin{tabular}{|c|c|c|}
\hline
$\rho_{\mbox{\tiny{min}}}/r_0$ & number of d.o.f. & $m_B r_0$ \\
\hline
$1.1$ & $11$ & $3.47(93)$ \\
\hline
$1.0$ & $14$ & $4.07(57)$ \\
\hline
$0.9$ & $17$ & $3.16(33)$ \\
\hline
$0.8$ & $18$ & $3.38(25)$ \\
\hline
$0.7$ & $21$ & $3.30(15)$ \\
\hline
$0.6$ & $23$ & $3.26(11)$ \\
\hline
$0.5$ & $26$ & $3.20(12)$ \\
\hline
$0.4$ & $29$ & $3.24(9)$ \\
\hline
\end{tabular}
\end{center}
\caption{Results for the dual gauge boson mass $m_B$, from the fit of eq.~(\ref{exdualsuperconductor}) to the numerical data, for different values of the lower cutoff on distances $\rho_{\mbox{\tiny{min}}}/r_0$.}
\label{mbfittab}
\end{table}
the data corresponding to distances shorter than a minimum value $\rho_{\mbox{\tiny{min}}}$; tab.~\ref{mbfittab} shows the results of the fit, for different values of $\rho_{\mbox{\tiny{min}}}$. Data analysis gives hints that a reliable choice for the lower distance cutoff to extract the dual gauge boson mass must be at least $\rho_{\mbox{\tiny{min}}} \simeq 0.8 r_0$: the corresponding curve is also shown in fig.~\ref{newfitfig}. The results obtained with this choice appear to be in quantitative agreement (within two errorbars) with the ones published in the literature.

A comment is in order: as opposite to the measurements of the interquark force at large distances, in the study of the electric field profile our numerical simulations of the dual model did not account for particular advantages with respect to state-of-the-art numerical studies in the standard formulation of compact QED: thus in this case it seems that to improve the result precision it is necessary to resort to larger statistics.

\section{Conclusions}\label{conclusionssect}

In this work, we addressed a numerical study of compact U(1) LGT in $D=4$, a theory which possesses a confining phase analogous to non-abelian models. We restricted our attention to the infinitely heavy quark limit, and studied the behaviour of a confined $Q\bar{Q}$ system, focusing our attention onto the interquark potential and force, and on the longitudinal component of the electric field induced in the mid-plane between the static charges.

We discussed the analytical transformation mapping the original system to a dual model formulated in terms of integer-valued variables, which opens the possibility to perform efficient simulations. In particular, this method allows to overcome the exponential growth of the relative error at large interquark distances (which affects measurements of the interquark potential and force in the standard approach to LGT's), focusing onto ratios of Polyakov loop correlation functions. To our knowledge, this is the first time this technique has been used to study these observables in compact U(1) LGT in $D=4$.

The numerical results that we obtained confirm the exponential error reduction with respect to Monte Carlo simulations in the conventional setting, and thus the method can be considered as a possible alternative to state-of-the-art error reduction algorithms, at least for the class of models for which this formulation is feasible. We measured data in a range of parameters which includes and generalises the results obtained with modern techniques, finding agreement with the data available in the literature. Then we compared the numerical results with the theoretical predictions. For the interquark potential and force, we compared our data with the expectations derived assuming that an effective string picture holds in the IR regime: we summarised the assumptions underlying this scenario, discussing both the theoretical issues (including the aspects that have been pointed out most recently) and the problems that are still open in the comparison with numerical results. Our findings are in agreement with the conclusions of analogous recent studies, confirming the validity of the effective string picture at large distances. We also extended this study to the finite temperature setting, presenting quantitative results for the string tension in a range of temperatures up to about $3/4$ of the critical deconfinement temperature.

Finally, we addressed the study of the longitudinal component of the electric field induced in the mid-plane between the static sources, comparing our numerical data with the prediction from the dual superconductor scenario, and measuring the dual gauge boson mass which, in the confined phase of the theory, drives the exponential-like decay of the electric field at large distance from the interquark axis.

\vskip1.0cm {\bf
Acknowledgements.}

It is a pleasure to thank Michele Caselle, who inspired the idea of this work, Burkhard Bunk, Yoshiaki Koma, Hendryk Pfeiffer, the organizers and the participants to the Lattice Gauge Theory meeting held in Turin (Italy) in December 2004 for enlightening discussions. The author acknowledges support received from Enterprise Ireland under the Basic Research Programme.

\end{document}